\documentclass{article}

\usepackage{arxiv}

\usepackage[utf8]{inputenc} % allow utf-8 input
\usepackage[T1]{fontenc}    % use 8-bit T1 fonts
\usepackage{hyperref}       % hyperlinks
\usepackage{url}            % simple URL typesetting
\usepackage{booktabs}       % professional-quality tables
\usepackage{amsfonts}       % blackboard math symbols
\usepackage{nicefrac}       % compact symbols for 1/2, etc.
\usepackage{microtype}      % microtypography
\usepackage{lipsum}
\usepackage{color}
\usepackage{algorithm}
\usepackage{algorithmic}
\usepackage{amsmath,amsfonts,amssymb}
\usepackage{subfigure}
\usepackage{lineno}
\usepackage{graphicx}
\usepackage{bm}
\usepackage{subfigure}
\graphicspath{{./figs/}}

\title{Compact representation of transonic airfoil buffet flows with observable-augmented machine learning}

\author{
Kai Fukami$^{*}$,
Yuta Iwatani,
Soju Maejima,
Hiroyuki Asada, and Soshi Kawai
\\
\\
Department of Aerospace Engineering, Graduate School of Engineering,\\
Tohoku University, Sendai, 980-8579, Japan\\
$^{*}$Corresponding author: kfukami1@tohoku.ac.jp
}

\begin{document}
\maketitle

\begin{abstract}

\vspace{-2mm}
{Transonic buffet presents time-dependent aerodynamic characteristics associated with shock, turbulent boundary layer, and their interactions.
Despite strong nonlinearities and a large degree of freedom, there exists a dominant dynamic pattern of a buffet cycle, suggesting the low dimensionality of transonic buffet phenomena.
This study seeks a low-dimensional representation of transonic airfoil buffet at a high Reynolds number with machine learning.}
Wall-modeled large-eddy simulations of flow over the OAT15A supercritical airfoil at two Mach numbers, $M_\infty = 0.715$ and 0.730, respectively producing non-buffet and buffet conditions, at a chord-based Reynolds number of $Re = 3\times 10^6$ are performed to generate the present datasets.
We find that the low-{dimensional} nature of transonic airfoil buffet can be extracted as a sole three-dimensional latent representation through lift-augmented autoencoder compression.
The current low-order representation not only describes the shock movement but also captures the moment when the separation occurs near the trailing edge in a low-order manner.
We further show that it is possible to perform sensor-based reconstruction through the present low-{dimensional} expression while identifying {the sensitivity with respect to aerodynamic responses.}
The present model trained at $Re = 3\times 10^6$ is lastly evaluated at the level of a real aircraft operation of $Re = 3\times 10^7$, exhibiting that the phase dynamics of lift is reasonably estimated from sparse sensors.
The current study may provide a foundation toward data-driven real-time analysis of transonic buffet conditions under aircraft operation.
\end{abstract}

\section{Introduction}
\label{sec:intro}

Transonic buffet phenomena determine the high-speed limit of flight envelope.
{To extend the flight envelope towards the high-speed side with a better and safer design of modern commercial aircraft that equips asymmetric supercritical wings, the transonic buffet, specifically referred to as Type II buffet~\cite{giannelis2017review}, needs to be tamed, rooted in profound understandings of its physics and practically useful models to describe the buffet.}
While extensive analyses using simulations and experiments have been performed providing a variety of posits to describe complex behaviors of transonic airfoil buffet,
what is commonly believed is that there exists a self-sustained shock buffet cycle~\cite{giannelis2017review}.
We pose a question of whether such a seemingly complex, but cyclic dynamics of transonic buffet phenomena can be described in a low-order manner with nonlinear machine learning.

{The aerodynamic instability known as transonic buffet, characterized by self-sustained shock wave oscillations on aircraft wings, needs to be taken into account during transonic or high-subsonic flight. 
This phenomenon arises because shock waves can form when the wing geometry accelerates the flow along the leading edge of the suction side, generating a localized supersonic region~\cite{tijdeman1980transonic}.}
The occurrence of transonic buffet depends on a flow condition characterized by a combination of parameters such as Mach number, Reynolds number, and angle of attack.

To facilitate characterizing the transonic buffet phenomena, a range of numerical and experimental endeavors have been carried out.
Such studies on the transonic buffet are classified based on their focus on dimension in phenomena, namely, two-dimensional and three-dimensional. 
{In the two-dimensional airfoil buffet, chord-wise large-scale shock oscillations occur, which are numerically and experimentally reproduced by confining a flow field in a narrow spanwise domain~\cite{lusher2024effectAIAAJ}.
{The chord-wise shock oscillations result in a distinct spectral peak at a low frequency generally smaller than $0.1$, for example, the Strouhal number $St\approx0.06$ for the OAT15A supercritical airfoil \cite{deck2005numerical, jacquin2009experimental,fukushima2018wall,cuong2022largeAIAAJ}.}
}

{
On the other hand, the three-dimensional buffet is caused due to characteristics associated with the three-dimensionality of the wing, such as swept and taper effects.
One notable feature of the three-dimensional buffet, while absent in the two-dimensional buffet, is the occurrence of buffet cells \cite{iovnovich2015numericalAIAAJ}. 
The buffet cells refer to a cellular flow structure propagating outboard. 
A range of numerical \cite{ohmichi2018modal,tamaki2024wall} and experimental studies \cite{dandois2016experimental,sugioka2018experimental,sugioka2021characteristicExpFluid,meneveau2000scale,masini2020analysisJFM} have reported the occurrence of the buffet cell. 
}

{
It has widely been observed that the power spectrum density of the relevant quantities, such as the pressure coefficient fluctuation, {typically} presents a broadband spectrum peak ranging from a Strouhal number of 0.2 to 0.6 \cite{koike2016unsteadyAIAAPaper,dandois2016experimental}, ten times higher in frequency than the two-dimensional buffet counterpart, depending on the sweep angle \cite{plante2020similaritiesAIAAJ, sugioka2022experimentalExpFluid,lusher2025implicitJFM}.
Particularly considering a full-aircraft configuration of the NASA Common Research Model, understanding of the buffet cell structure has been deepened with modal analysis, including the tri-global stability analysis~\cite{timme2020globalJFM,sansica2023globalAIAAJ}, the tri-resolvent analysis~\cite{houtman2023resolventFlow}, dynamic mode decomposition~\cite{ohmichi2018modal}, and its Hankel variant~\cite{asada2025exact}.
Based on them, the buffet cell has been recognized as a key player in the self-sustaining instability mechanism of the three-dimensional buffet. 
However, there is still no widely accepted physical model that explains the self-sustaining mechanism of a three-dimensional buffet.
}

{
While acknowledging the significance of buffet cells, this study focuses on the two-dimensional airfoil buffet mechanism, which remains active and critical even under three-dimensional buffet conditions.
Sugioka et al.~\cite{sugioka2018experimental} experimentally demonstrated that shock wave oscillations over the NASA Common Research Model at high angles of attack exhibit behavior similar to a two-dimensional buffet.
Paladini et al.~\cite{paladini2019transonicPRF} showed that a two-dimensional global instability mode, akin to the one observed in airfoil buffet \cite{crouch2009origin}, can coexist with a spanwise-varying three-dimensional mode associated with buffet cells.
Similar modal structures have been reported by Crouch et al.~\cite{crouch2018globalFDconf,crouch2019globalFJM}.
Paladini et al.~\cite{paladini2019transonicPRF} performed a wavemaker analysis to reveal that the two-dimensional mode is primarily linked to the shock wave dynamics, whereas the spanwise-varying mode originates from the separated shear layer.
These findings highlight the importance of considering not only the three-dimensional buffet cells but also the underlying two-dimensional instability mechanisms that remain fundamental to understanding buffet phenomena.
}

{
For these reasons, the mechanism of self-sustained large-scale shock oscillations is of particular interest in the community ~\cite{lee2001self,iwatani2023identifying}.
}
While a Reynolds-averaged formulation had been considered for numerical investigations~\cite{crouch2009origin,iovnovich2012AIAAJ,sartor2015stability}, recent advancements in computational resources along with wall-modeling approaches enable performing large-eddy simulations~\cite{fukushima2018wall,tamaki2024wall,goc2025studies}.
This offers further reliable assessments of transonic buffet flows by accurately capturing the interaction between the shock wave and the turbulent boundary layer.
Along with spatiotemporal high-resolution measuring techniques such as laser Doppler velocimetry~\cite{jacquin2009experimental}, particle image velocimetry~\cite{d2021experimental}, and Schlieren visualization~\cite{schauerte2023experimentalAIAAJ}, experimental studies have not only provided a simplified model of transonic buffet supporting the understanding of buffet phenomena~\cite{lee1990oscillatory,crouch2007predicting} but also suggested passive control devices to suppress the buffet-associated instabilities~\cite{lagemann2024towards}.
However, the self-sustaining mechanisms of the transonic airfoil buffet still require further clarification.

In analyzing the transonic buffet flows with a large degree of freedom in the direction of space, time, and flow parameters, one can consider applying data-driven order-reduction techniques to flow field snapshots made available through simulations and experiments.
For example, proper orthogonal decomposition~\cite[POD;][]{Lumely1967} has been considered to obtain a low-order representation of transonic buffet phenomena~\cite{ohmichi2018modal,poplingher2019modal,sansica2022system,iwatani2022pod}.
However, seeking a minimal representation of unsteady flows with such a linear technique is generally challenging because given data are linearly projected onto a flat manifold~\cite{graham2021exact}.

To extract a low-order representation that best captures the underlying characteristics of transonic buffet flows from data, this study considers a nonlinear autoencoder-based compression~\cite{HS2006}.
Nonlinear activation functions inside autoencoder enable better compression of unsteady flow data compared to linear techniques, which has been discussed with wake shedding~\cite{omata2019novel,MFF2019}, channel flow~\cite{FNKF2019,yousif2022physics}, Kolmogorov turbulence~\cite{page2024exact}, and aerodynamic flows under gusty environments~\cite{ME2025}.
Compressed representations obtained from the autoencoder can be used for a range of analyses including mode decomposition~\cite{FNF2020,MTM2024}, dynamical modeling~\cite{fukami2020sparse,solera2024beta,constante2024data}, shape optimization~\cite{tran2024data}, and flow control~\cite{linot2023turbulence,liu2024model}.

Although nonlinear autoencoder can be employed as a powerful data compressor of unsteady flows, it is important to note that careful use of autoencoder by incorporating prior knowledge of physics is essential to promote understanding {of} flows in a low-order latent space~\cite{FT2023}.
It is challenging to use compressed variables obtained through a na\"ive application of a standard autoencoder for characterizing and controlling unsteady flows~\cite{fukami2024data,smith2024cyclic}.
In response, we incorporate aerodynamic coefficients into the nonlinear autoencoder formulation in identifying a low-order subspace.
Equipped with this observable-augmented autoencoder, this study unveils the existence of a three-dimensional representation of transonic airfoil buffet flows, which describes the complex phenomena over the buffet cycle dynamics in a compact manner.
Furthermore, the current model trained at a wind-tunnel scale Reynolds number based on a chord length $Re\sim 10^6$ can be used for sparse sensor reconstruction of aerodynamic responses at the level of a real-aircraft operation high Reynolds number $Re\sim 10^7$.
The present approach may facilitate data-driven analysis of transonic buffet flows across a range of Reynolds numbers.

This paper is organized as follows.
The simulation setup used for data generation and flow physics are expressed in section~\ref{sec:TABF}.
The present autoencoder technique is described in section~\ref{sec:AE}.
Results and discussion are presented in section~\ref{sec:res}. 
Conclusions are offered in section~\ref{sec:conc}.

% \vspace{-3mm}
\section{Transonic airfoil buffet flows at high Reynolds numbers}
\label{sec:TABF}

\begin{figure}
    \centering
    \includegraphics[width=\textwidth]{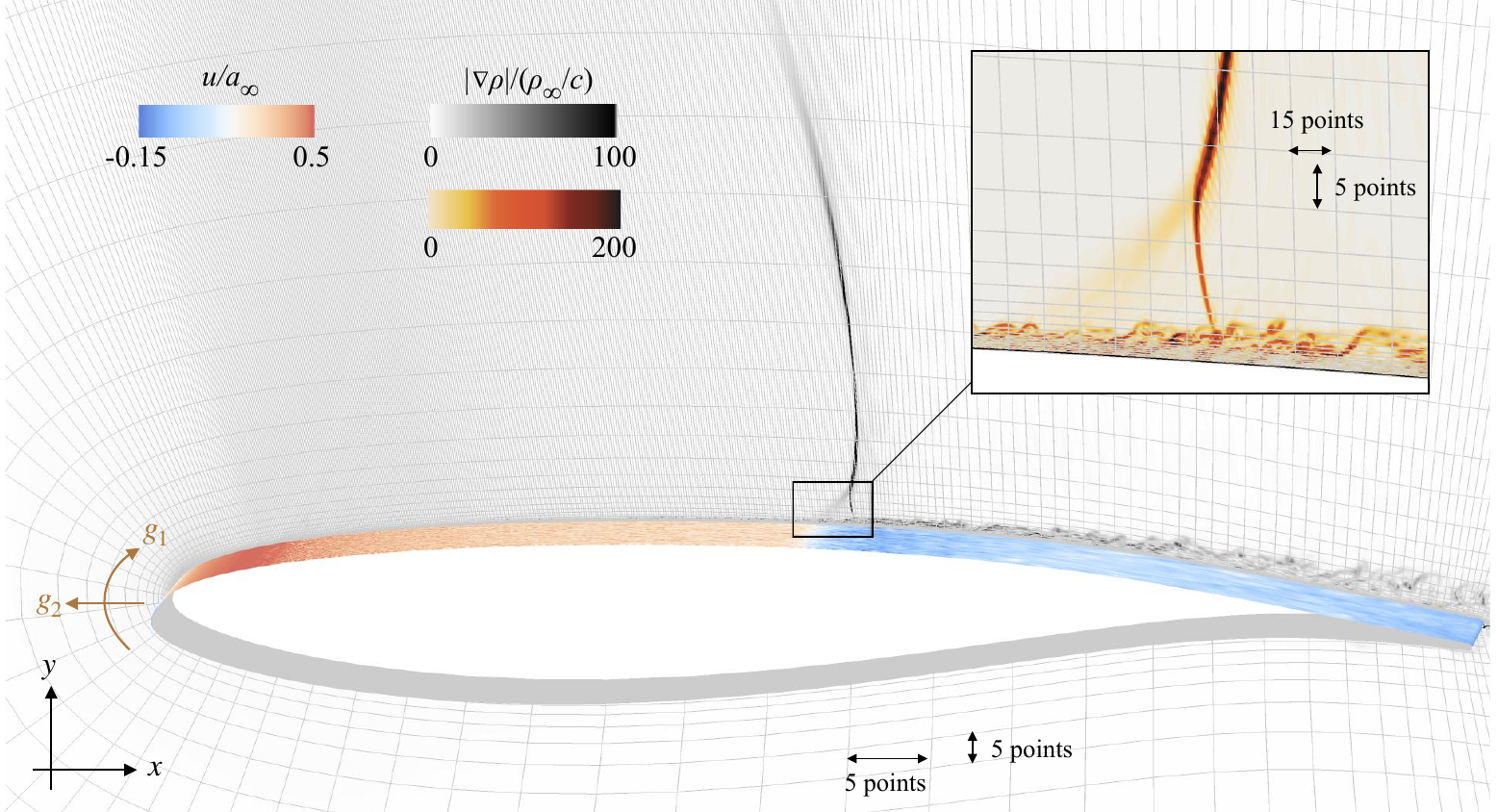}
    % \vspace{-5mm}
    \caption{
    The computational grid used in the present wall-modeled large-eddy simulations of two-dimensional transonic airfoil buffet at a high Reynolds number~\cite{fukushima2018wall}. 
    An instantaneous streamwise velocity field $u$ near the wall and the density gradient magnitude $|\nabla \rho|$ are superposed.     
    The gray grid lines are displayed every fifth point in the $g_1$ and $g_2$ (wall-normal) directions. 
    The subfigure is focused on the region of the shock wave-turbulent boundary layer interactions with the gray grid lines plotted every fifteenth point in the $g_1$ direction and every fifth point in the $g_2$ direction.
    }
    % \vspace{-3mm}
    \label{fig1}
\end{figure}

This study seeks a low-{dimensional} representation of two-dimensional transonic airfoil buffet flows, capturing time-varying characteristics over the buffet cycle using nonlinear machine learning.
We consider datasets of Fukushima and Kawai~\cite{fukushima2018wall} generated by wall-modeled large-eddy simulation (LES) of the transonic buffet over the OAT15A supercritical airfoil at a high Reynolds number of $Re = u_\infty c/\nu_\infty = 3\times 10^6$ for nonlinear machine-learning compression.
Here, $u_\infty$, $c$, and $\nu_\infty$ describe the free-stream velocity, the chord length, and the kinematic viscosity, respectively.
Following the observation in our previous study~\cite{fukushima2018wall}, we consider two different Mach numbers of $M_\infty = u_\infty/a_\infty = (0.715, 0.730)$, where $a_\infty$ is the freestream sonic speed.
While the steady shock wave is observed at $M_\infty = 0.715$, the unsteady shock oscillating buffet phenomena emerge by increasing the Mach number to 0.730.
Involving both non-buffet and buffet conditions in the present datasets for the nonlinear machine-learning analysis enables extracting the difference between them in a low-order manner.
All the physical variables throughout the paper are normalized using combinations of $c$, $a_\infty$, and the density $\rho_\infty$.
We further consider a higher Reynolds number case of $Re = 3\times 10^7$ with $M_\infty = 0.730$, exhibiting the unsteady buffet phenomena, to evaluate the applicability of the current technique trained at a wind tunnel-scale Reynolds number $Re \sim 10^6$ to a scenario at a real aircraft-scale Reynolds number $Re \sim 10^7$.

The computational mesh used in the present study is shown in figure~\ref{fig1}.
The spatially filtered compressible Navier--Stokes equations are numerically solved, where the LES with modeled wall shear stresses and wall heat fluxes resolves the outer-layer turbulence~\cite{fukushima2018wall}.
We follow our previous studies~\cite{fukushima2018wall,kawai2012wall,kawai2013dynamic} for the numerical schemes as well as the treatment of the boundary conditions.

\begin{figure}
    \centering
    \hspace{-8mm}
    \includegraphics[width=1.05\textwidth]{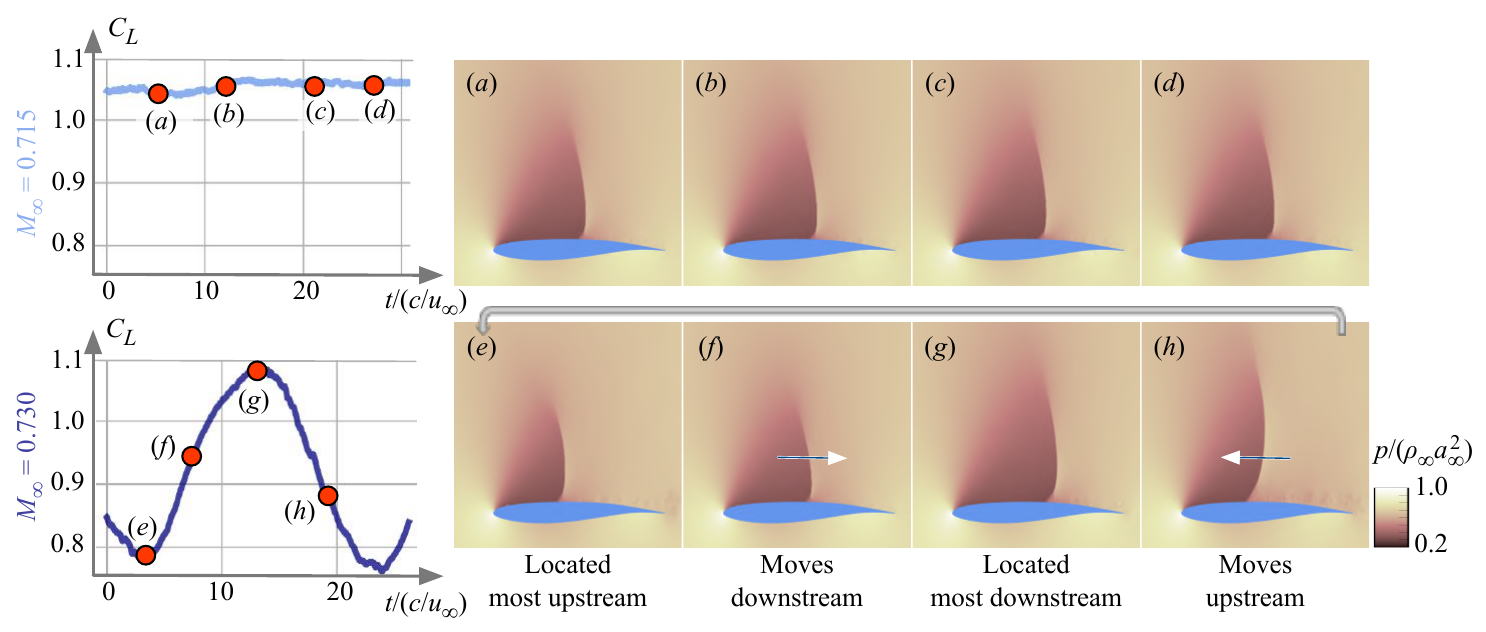}
    % \vspace{-5mm}
    \caption{
    Lift coefficient and pressure fields at $M_\infty=0.715$ (top) and 0.730 (bottom).
    A note for the shock location is provided underneath each contour of $M_\infty = 0.730$.
    The arrow in each subcontour represents the direction of shock movement.
    }
    % \vspace{-3mm}
    \label{fig2}
\end{figure}

The spatial derivatives at interior grid points are evaluated using the sixth-order compact differencing scheme~\cite{lele1992compact}.
Time integration is performed with the third-order total variation diminishing Runge--Kutta scheme~\cite{gottlieb1998total}.
To accurately resolve the shock wave, the localized artificial diffusivity method is employed with the sixth-order compact scheme~\cite{kawai2010assessment}.
While we compute the subgrid-scale turbulent eddy viscosity with a selective mixed-scale model~\cite{lenormand2000large}, the equilibrium wall model~\cite{kawai2012wall} is considered.

{
The computational mesh for the present wall-modeled LES is designed based on the grid resolution requirements \cite{kawai2012wall, larsson2016largeJSME}.  
Although we use the same mesh at both Reynolds numbers at $Re = 3\times 10^6$ and $3\times 10^7$, the employed mesh satisfies the resolution requirements across the streamwise domain of the attached fully turbulent boundary layer upstream of the shock wave ($0.2\lesssim x/c \lesssim 0.35$), providing more than 23-25 grid points in each direction per boundary layer thickness. 
Specifically, the mesh resolves the boundary layer with at least 29, 34, and 38 points in the wall-normal direction at $x/c \approx 0.2$, 0.25, and 0.3, respectively.
In the wall-parallel directions, the resolution corresponds to at least 23, 28, and 33 grid points per local boundary layer thickness at the same stream locations.
These values meet the standards for wall-modeled LES resolution~\cite{kawai2012wall}.
}

Furthermore, previous studies have reported that the wall-modeled LES with the equilibrium wall model can reasonably produce the flow states associated with the interaction between the shock waves and turbulent boundary layer even with the simplification of the equilibrium wall model~\cite{bermejo2014confinement,fukushima2018wall,de2022wallIJHFF,sashida2024wallAIAApaper}.
{Therefore, the present wall-modeled LES provides a high-fidelity dataset for the present nonlinear machine-learning analysis.}
Further details on the simulation setup are referred to Fukushima and Kawai~\cite{fukushima2018wall}.

The temporal evolution of lift coefficient $C_L$ and a sectional pressure field $p$ extracted from the wing center in the spanwise direction at $Re=3\times 10^6$ obtained through the present simulation is presented in figure~\ref{fig2}.
The case for $M_\infty = 0.715$ shows statistically steady states, producing small fluctuations of lift over time.
The shock mostly appears at $x/c\approx 0.55$ while slightly oscillating in the streamwise direction on the wing.

In contrast, the case for $M_\infty = 0.730$ clearly presents its time-varying feature associated with self-sustained large-scale shock oscillation.
The shock wave periodically moves in large amplitude over a wing while the separation near the trailing edge is triggered depending on the shock location, which coincides with observations in wind-tunnel experiments~\cite{jacquin2009experimental}.
Correspondingly, the lift response also exhibits a periodic signal over the buffet cycle.
Hence, the phase of shock location over the buffet cycle is almost identical to that of lift.
The separation height is particularly increased when the shock wave moves upstream, which will be shown later.
The interaction between the wake and separation at this stage causes the upstream traveling wave~\cite{lee2001self,d2021experimental,iwatani2023identifying}.
The lift response is greatly affected by the time-varying area size of supersonic flow along with the aforementioned processes.
Note that these buffet dynamics are further discussed and quantified later with the observation in the machine-learning-based low-{dimensional} subspace.

% \vspace{-3mm}
\section{Nonlinear machine-learning-based compression of transonic airfoil buffet flows}
\label{sec:AE}

\begin{figure}
    \centering
    \includegraphics[width=0.95\textwidth]{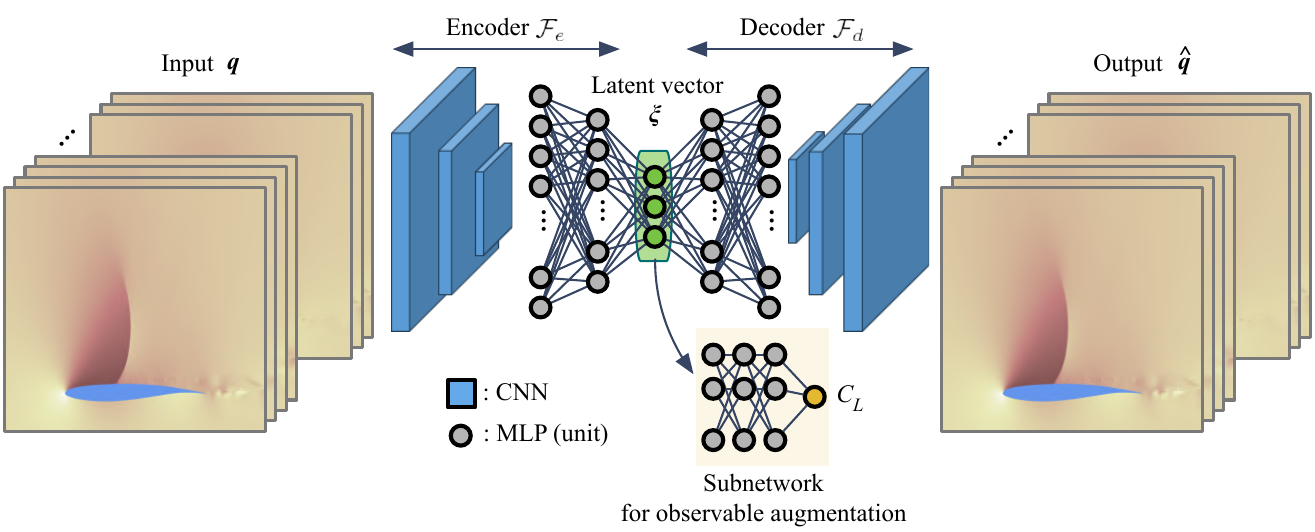}
    % \vspace{-5mm}
    \caption{
    Lift-augmented nonlinear autoencoder~\cite{FT2023}.
    }
    % \vspace{-3mm}
    \label{fig3}
\end{figure}

%% number of snapshot

To seek a low-{dimensional} representation of transonic airfoil buffet flows from data, we consider a nonlinear autoencoder-based data compression~\cite{HS2006}.
An autoencoder ${\cal F}_{\rm AE}$ aims to reconstruct (or output) the same data as the input data ${\bm q}\in \mathbb{R}^n$.
The autoencoder is designed to possess the bottleneck, referred to as a latent space ${\bm \xi} \in \mathbb{R}^{m}$, as illustrated in figure~\ref{fig3}.
The latent dimension $m$ is generally set to be much smaller than the original data dimension $n$ such that $m\ll n$.
Hence, the latent vector ${\bm \xi}$ can be considered as a compressed representation of the given data ${\bm q}$ if the autoencoder ${\cal F}_{\rm AE}$ accurately reconstructs the data.
The aforementioned process is described as
\begin{align}
    {\bm q}\approx{\cal F}_{\rm AE}(\bm q) = {\cal F}_d({\cal F}_e({\bm q})),
    ~~~
    {\bm \xi} = {\cal F}_e(\bm q),~~~ {\bm q} \approx \widehat{\bm q} = {\cal F}_d({\bm \xi}),
\end{align}
where $\widehat{(\cdot)}$ denotes a reconstructed variable, and ${\cal F}_e$ and ${\cal F}_d$ correspond to an encoder and a decoder, respectively.
A range of neural-network models with nonlinear activation functions can be considered for the construction of autoencoder ${\cal F}_{\rm AE}$.
The use of nonlinear activation functions promotes network capabilities, providing better compression than linear techniques, which is mathematically proven through the relationship between a linear-activation autoencoder and other linear compression approaches~\cite{oja1982simplified,bourlard1988auto,fukami2021model}.

\begin{table}
\begin{center}
\begin{tabular}{cc|cc|cc}
\multicolumn{2}{c|}{Encoder}          & \multicolumn{2}{c|}{Decoder}        & \multicolumn{2}{c}{Lift subnetwork} \\ \cline{1-6}
Layer                 & Data size     & Layer              & Data size      & Layer                & Data size   \\ \cline{1-6}
Input $\bm{q}$ & (480, 200) & Latent vector $\bm{\xi}$ & (3)            & Latent vector $\bm{\xi}$   & (3)         \\
Conv. (3, 3, 16)       & (480, 200, 16) & MLP                & (16)            & MLP                  & (32)        \\
Conv. (3, 3, 16)       & (480, 200, 16) & MLP                & (32)           & MLP                  & (64)        \\
Maxpooling (2, 2)     & (240, 100, 16)   & MLP                & (256)           & MLP                  & (32)        \\
Conv. (3, 3, 16)      & (240, 100, 16)  & MLP                & (480)          & Output $\widehat{C}_L$& (1)         \\
Conv. (3, 3, 16)      & (240, 100, 16)  & (Reshape)          & (12, 5, 8)         &                   &         \\
Maxpooling (2, 2)     & (120, 50, 16)  & Conv. (3, 3, 8)        & (12, 5, 8)        &   &          \\
Conv. (3, 3, 8)      & (120, 50, 8)  & Conv. (3, 3, 8)          & (12, 5, 8)   &                      &             \\
Conv. (3, 3, 8)      & (120, 50, 8)   & Upsampling (5, 5)   & (60, 25, 8)    &                      &             \\
Maxpooling (2, 2)     & (60, 25, 8)  & Conv. (3, 3, 8)      & (60, 25, 8)        &   \\
Conv. (3, 3, 8)      & (60, 25, 8)  & Conv. (3, 3, 8)      & (60, 25, 8)    &                      &             \\
Conv. (3, 3, 8)      & (60, 25, 8)   & Upsampling (2, 2)   & (120, 50, 8)    &                      &             \\
Maxpooling (5, 5)     & (12, 5, 8)  & Conv. (3, 3, 16)     & (120, 50, 16)       &   \\
Conv. (3, 3, 8)      & (12, 5, 8)  & Conv. (3, 3, 16)     & (120, 50, 16)   &                      &             \\
Conv. (3, 3, 8)      & (12, 5, 8)   & Upsampling (2, 2)    & (240, 100, 16)    &                      &             \\

(Reshape)             & (480)        & Conv. (3, 3, 16)   & (240, 100, 16)   &                      &             \\
MLP                   & (256)         & Conv. (3, 3, 16)   & (240, 100, 16)   &                      &             \\
MLP                   & (64)         & Upsampling (2, 2)   & (480, 200, 16)   &                      &             \\
MLP                   & (32)          & Conv. (3, 3, 16)   & (480, 200, 16)   &                      &             \\
MLP                   & (16)          & Conv. (3, 3, 16)  & (480, 200, 16) &                      &             \\
Latent vector $\bm{\xi}$
& (3)           &  Output $\widehat{\bm{q}}$  & (480, 200) &                      &             
\end{tabular}
\caption{
The architecture of observable-augmented nonlinear autoencoder. 
The convolutional layers are denoted as ``Conv." 
The size of the convolutional filter $F$ and the number of the filter $K$ are shown for each convolutional layer as $(F, F, K)$.
The maxpooling/upsampling ratio $R$ is shown for each layer as $(R, R)$.
}
\label{tab:aemodel}
\end{center}
\end{table}

We consider a sectional pressure field sampled from the wing center in the spanwise direction as the input and output ${\bm q}$ of a nonlinear autoencoder to extract the underlying characteristics of transonic airfoil buffet flows.
While a standard autoencoder achieves significant data compression of fluid flows, it is often challenging to interpret the identified subspace in a physically understandable manner.
To facilitate the present latent identification from the viewpoint of aerodynamics, this study uses a lift-augmented nonlinear autoencoder~\cite{FT2023} producing a lift response from the latent vector through a branch network, as illustrated in figure~\ref{fig3}.
The optimization for the parameters (or weights) $\bm w$ inside the lift-augmented autoencoder is performed with
{
\begin{align}
    {\bm w}^* = {\rm argmin}_{\bm w}\biggl[||{\bm q}-\hat{\bm q}||^2_2 + \beta ||C_L - \hat{C_L}||^2_2 \biggr],
    \label{eq:obsloss}
\end{align}}
where $\beta$ balances the pressure field and lift reconstruction loss terms.
This weighting parameter $\beta$ is set to 0.03 and 0.05 based on the L-curve analysis~\cite{hansen1993use} {for the observable-augmented autoencoder, while a regular autoencoder, i.e., $\beta = 0$ is also considered for comparison.}
To minimize the above cost function, the model needs to accurately estimate $C_L(t)$ while performing data compression of the pressure field data ${\bm q}(t)$.
In other words, the current formulation enables ${\bm w}$ {to be tuned to capture structures appearing over the buffet cycle that are associated with the lift response.
As the periodic shock movement over an OAT15A airfoil, clearly observed in the pressure field, is highly correlated with the lift coefficient $C_L(t)$, the resulting low-{dimensional} representation is expected to emphasize aerodynamically important events during the buffet cycle.}

The current data set for the nonlinear autoencoder analysis is composed of 6,800 snapshots with $M_\infty = 0.715$ (non-buffet condition) over 30.8 non-dimensional time, $t/(c/u_\infty)$, and 17,300 snapshots with $M_\infty = 0.730$ (buffet condition) over 26.4 non-dimensional time.
We consider a subdomain of $(x, y)/c \in [-0.6, 1.5] \times [-0.5, 1.3]$ with spatially uniform grid points $(N_x, N_y) = (480, 200)$ extracted from the entire computational domain for the data-driven analysis, where the leading edge of the wing is positioned at the origin. 
{The interior of the wing is set to be zero.
As a fixed angle of attack is considered for all the data in this study, the model is not affected by this operation.}
The present autoencoder is composed of convolutional neural networks~\cite{LBBH1998} and multi-layer perceptrons~\cite{RHW1986} following the original study of the lift-augmented autoencoder, as summarized in table~\ref{tab:aemodel}.
{While the convolutional network learns large-scale structures in a flow field through filter-based operations, the multi-layer perceptrons are used for the bottleneck part of the autoencoder, where the data dimension is very low and the spatial coherence is less important than the complex relationship among the latent variables~\cite{fukagata2025compressing}.
This combination enables data-driven compression of fluid flow data with reasonable computational costs compared to a model based solely on a multi-layer perceptron that often encounters the curse of dimensionality~\cite{fukami2021model,morimoto2021convolutional}.
}
Further details on machine-learning setups {with the present L-curve analysis for the decision of $\beta$ are given in Appendix A} and a sample code (\url{https://github.com/kfukami/Observable-AE}).

% \vspace{-3mm}
\section{Results and discussion}
\label{sec:res}
% \vspace{-10mm}

\subsection{Latent space identification of transonic airfoil buffet flows}

\begin{figure}
    \centering
    \includegraphics[width=0.85\textwidth]{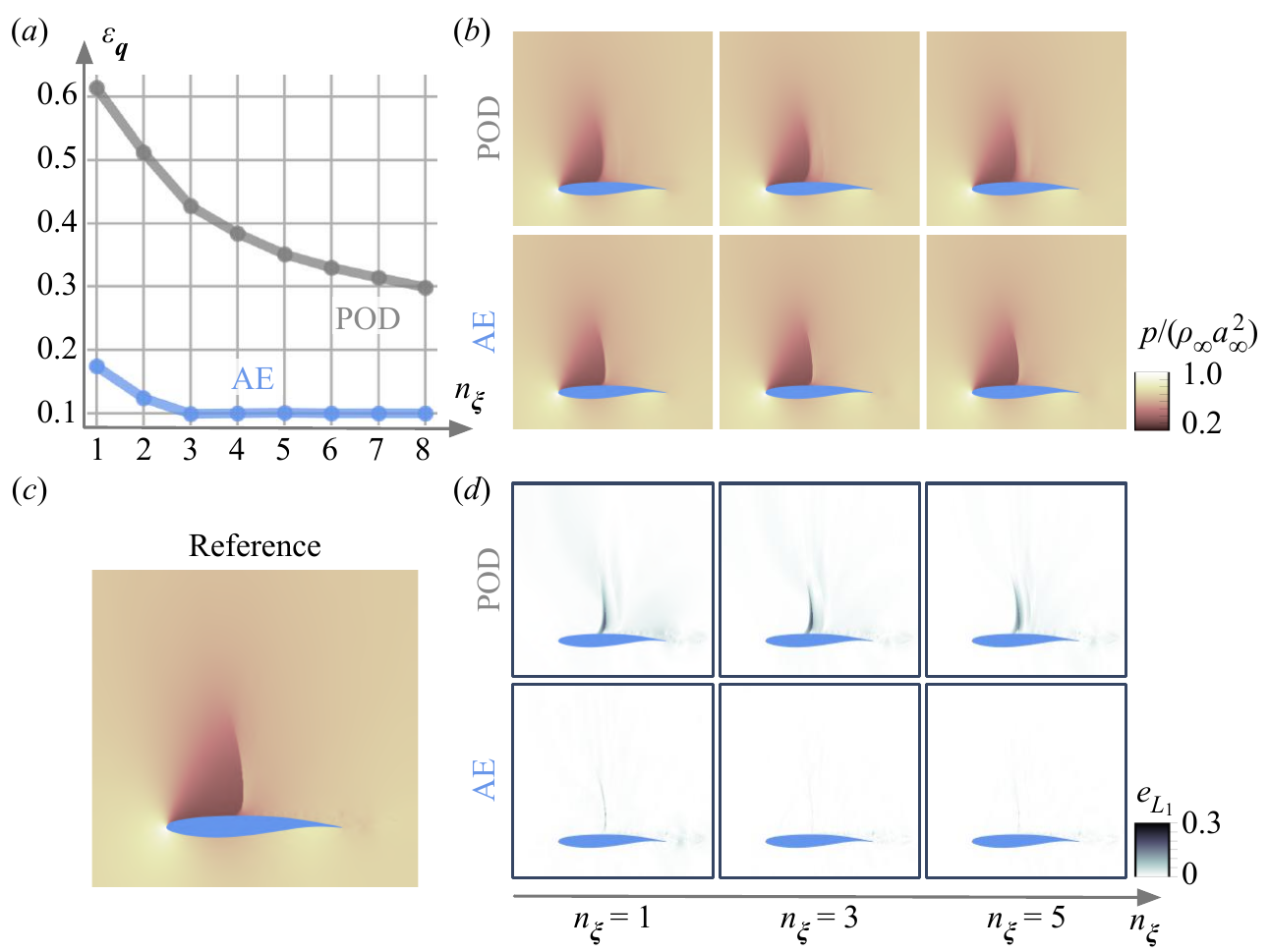}
    % \vspace{-5mm}
    \caption{
    {
    Comparison of compression performance for transonic airfoil buffet flow data between linear POD and a standard nonlinear autoencoder (AE, $\beta=0$).
    $(a)$ The relationship between the latent dimension $n_{\bm \xi}$ and the $L_2$ reconstruction error $\varepsilon$. 
    $(b)$~Representative reconstructed pressure snapshots with $n_{\bm \xi} = (1,3,5)$ for $M_\infty = 0.730$ with $(c)$ the reference field.
    $(d)$ The absolute error field $e_{L_1} = |{\bm q}-\hat{\bm q}|$ corresponding to figures~$(b)$.}
    }
    % \vspace{-3mm}
    \label{fig4}
\end{figure}

This section discusses data-driven compression and the resulting subspace identification of the transonic airfoil buffet flows.
Let us first examine the latent dimension that accurately reproduces the original flow state.
The relationship between the latent dimension $n_{\bm \xi}$ and the $L_2$ reconstruction error norm ${\varepsilon}_{\bm q}$ is shown in figure~\ref{fig4}.
Here, the $L_2$ reconstruction error norm between a variable ${\bm f}$ and its reconstruction $\hat{\bm f}$ is defined as $\varepsilon_{\bm f} = ||{\bm f}-\hat{\bm f}||^2_2/||{\bm f}^\prime||^2_2$, where ${\bm f}^\prime$ represents the fluctuation of ${\bm f}$ from the time-averaged value.
{While a standard nonlinear autoencoder without lift incorporation, i.e., $\beta=0$, is considered for this analysis}, the linear POD is also used for comparison.

The nonlinear autoencoder is superior to POD across the latent dimension, suggesting that the use of nonlinear activation functions inside the model facilitates compression performance.
Compared to the POD-based reconstruction exhibiting high error near the shock, the autoencoder accurately reproduces a flow state, as presented in figure~\ref{fig4}.
We also find that the error curve of autoencoder plateaus once the latent dimension reaches three.
This reveals that the primary {large-scale} feature of the pressure fields for the present transonic airfoil buffet flows at $Re=3\times 10^6$ can be represented with solely three-dimensional latent variables with nonlinear machine learning.
{To achieve a similar reconstruction level of $\varepsilon_{\bm q} \approx 0.1$ to a nonlinear autoencoder with $n_{\bm \xi}=3$, 85 linear POD modes are needed.}

{The plateau behavior for the autoencoder is in part due to the present network architecture shown in table~\ref{tab:aemodel}, which compresses data with 480 dimensions given by the portion of the convolutional network to be ${\cal O}(10^0)$ using multi-layer perceptrons.
A similar observation of producing plateau behavior in capturing dominant large-scale features has recently been found in Fukami et al.~\cite{FST2025} for extremely strong vortex-airfoil interactions with turbulent vortical structures.
It is anticipated that the error would be further reduced once fine-scale structures begin to be captured in the latent space with much larger latent dimensions.
Since large-scale motions have already been extracted with $n_{\bm \xi} = 3$, the resulting curve for the autoencoder likely exhibits a step-type behavior in which the plateaued error reduces again once the latent dimension becomes sufficiently large.}
Hereafter, we choose the latent dimension of 3 for the discussions.

\begin{figure}[t]
    \centering
    \includegraphics[width=0.9\textwidth]{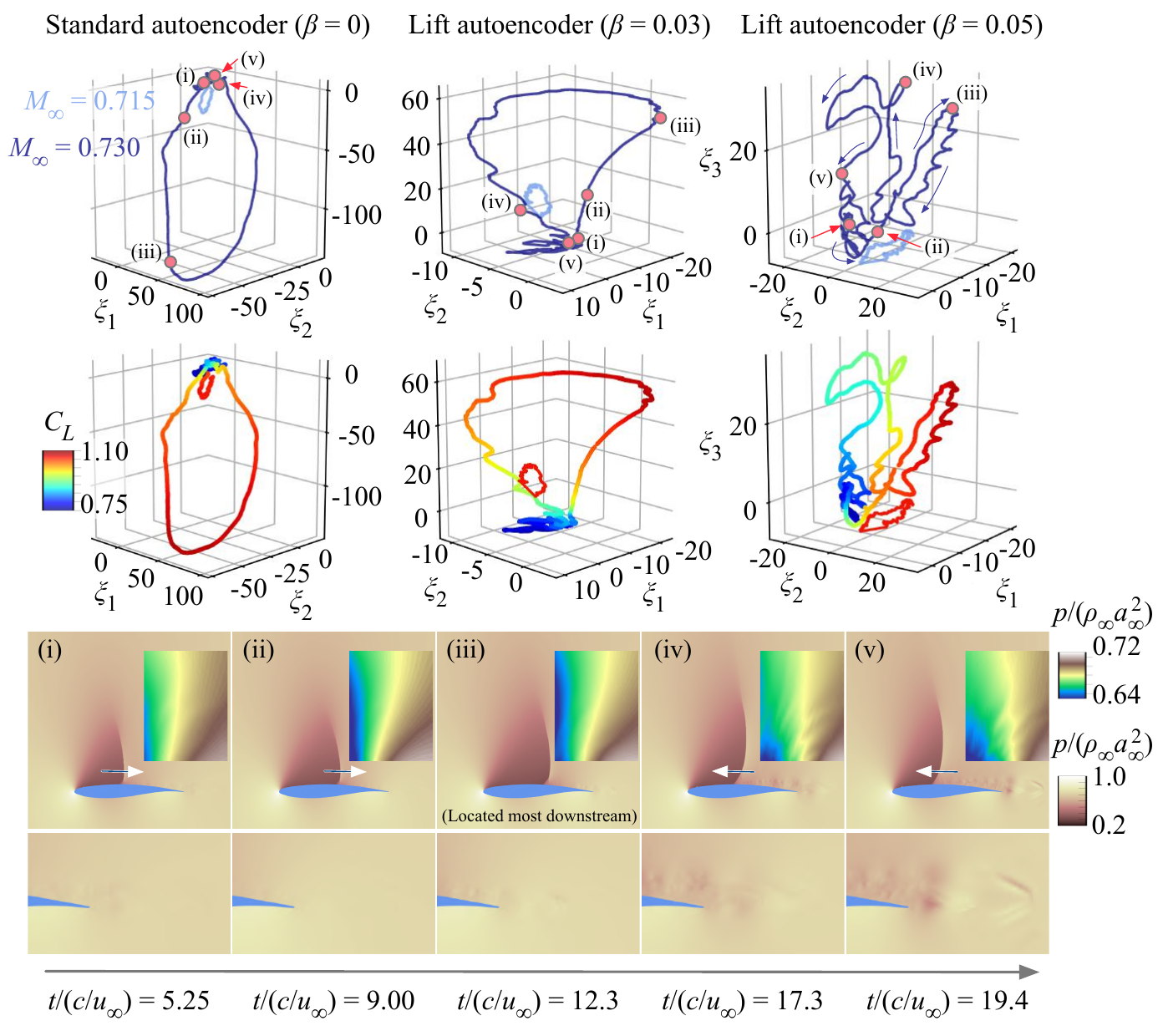}
    % \vspace{-5mm}
    \caption{
    Latent subspace identified by a standard autoencoder ($\beta=0$) and the lift-augmented autoencoder ($\beta=0.03$ and 0.05) colored by the cases of different Mach numbers $M_\infty = (0.715, 0.730)$ (top) and the time-varying lift coefficient $C_L(t)$ (bottom).
    The pressure fields over time corresponding to the points $\rm (i-iv)$ in the latent space are also shown.
    The arrow in each subcontour represents the direction of shock movement.
    The zoomed-in view of wake and the downstream region visualized with a different color scheme are also depicted to emphasize the interaction between the wake, shock, and turbulent boundary layer.
    }
    \vspace{-3mm}
    \label{fig5}
\end{figure}

Next, we examine the behavior of low-dimensionalized transonic airfoil buffet flows in the latent space.
The three-dimensional subspace identified by a standard autoencoder ($\beta=0$) and the lift-augmented autoencoder ($\beta=0.03$ and 0.05) is exhibited in figure~\ref{fig5}.
For all the cases, the trajectory for the non-buffet and buffet cases appears in different regions of the latent space.
The non-buffet case for $M_\infty = 0.715$ across the autoencoders is described in a similar way, that is, a small-sized circle-like orbit.
This representation likely corresponds to the statistically steady dynamics with small oscillations of aerodynamic responses for the present non-buffet flows{, which is evident from the reconstruction of lift response and pressure fields for the non-buffet case presented in Appendix B.}

While all the present subspaces capture the relationship between the non-buffet and buffet cases and the characteristics of the non-buffet flow in a low-order manner, the latent expression for the buffet case of $M_\infty = 0.730$ shows a clear difference by introducing the lift augmentation.
This can be observed with the difference in the relative location of the low-dimensionalized flow states $\rm (i)$, $\rm (iv)$, and $\rm (v)$.
Here, the shock in the flow field $\rm (i)$ moves downstream while that in $\rm (iv)$ and $\rm (v)$ moves upstream.
The standard model encodes them into nearby regions in the latent space.
In contrast, their locations begin to differ due to the lift augmentation.
Consequently, the low-order trajectory with $\beta = 0.05$ presents a geometric structure possessing two wings, while that with $\beta =0$ and 0.03 rather shows a regular cyclic orbit.

% Then 0.735

\begin{figure}
    \centering
    \includegraphics[width=0.9\textwidth]{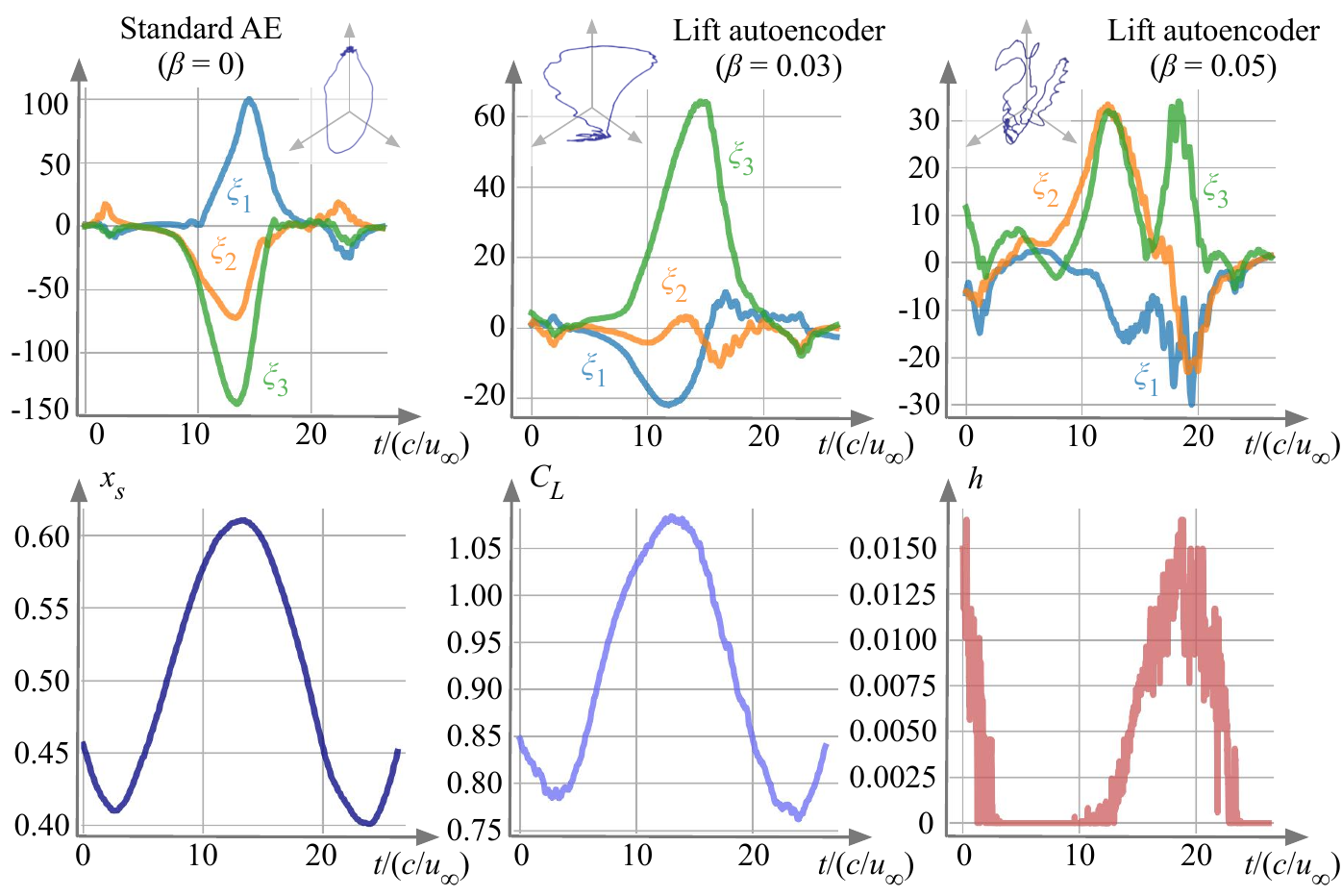}
    % \vspace{-5mm}
    \caption{
    Time trace of latent vectors ${\bm \xi}$ obtained from nonlinear autoencoders, shock location $x_s$, lift coefficient $C_L$, and separation height $h$ for the buffet case.
    }
    % \vspace{-6mm}
    \label{fig6}
\end{figure}

To discuss what physics are captured in the present low-order representation, the temporal behavior of latent vectors ${\bm \xi}(t)$ is compared to the shock location $x_s(t)$, the lift coefficient $C_L(t)$, and the separation height $h(t)$, as shown in figure~\ref{fig6}.
Here, the shock location $x_s$ is defined as a streamwise position at which the density gradient magnitude $|\nabla \rho|$ takes the maximum value.
The separation height $h$ is set to be a distance from the wall in which the streamwise momentum $\rho u$ becomes 0 at $x/c = 0.6$ in measuring across the wall-normal direction.

The latent expression from the standard autoencoder emphasizes the cyclic behavior of shock location as the notable peak of latent vectors at $t\approx 13$.
With the lift incorporation of $\beta = 0.05$, the latent vectors possess the additional dominant peak around $t=20$, corresponding to the emergence of the wing-type geometric structure in the low-order subspace.
While this moment is under-evaluated with $\beta = 0$ and 0.03, we find that the appearing peak at $t\approx 20$ coincides with the timing when the separation height $h$ is increased, as shown in figure~\ref{fig6}.
This increase in the separation height $h$ is attributed to the upstream moving shock wave, not only producing a strong shock due to the increase of relative shock Mack number but also inducing a large separation due to a strong shock adverse pressure gradient.
In this manner, the separation height varies depending on the direction of shock movement across the streamwise direction, i.e., relative shock Mach number, in addition to the shock location.
Hence, it can be argued that the current lift augmentation well captures the relationship between the shock motion and the aerodynamic responses in its latent representation.
Although the flow field data itself given as the input may also include the phase information of buffet cycle as the phase of shock location matches that of lift response as presented in figure~\ref{fig6}, the present observation suggests that providing an aerodynamic variable as an observable output through the subnetwork is essential to identify the physically interpretable subspace.
{The dependence of the latent representation geometry on the number of training samples and the initial random seed assigned to the weights in the observable-augmented autoencoder is examined in Appendices C and D, respectively.}

{
Note that all the latent spaces across $\beta$ represent the cyclic transonic buffet dynamics while achieving the same level of reconstruction through the decoder.
The latent expression hence becomes stretched by highlighting the events associated with a given observable.
In other words, all the latent subspaces are regarded as the compact representation of transonic airfoil buffet flows, although their presentation ways are different from each other.
The present lift augmentation can highlight aerodynamically important events as a manifold geometry while a regular model does not capture them in an interpretable manner, e.g., points (i), (v), and (iv) in figure~\ref{fig5}.
}

\subsection{Sparse sensor reconstruction of transonic airfoil buffet flows via low-order subspace}

\begin{figure}
    \centering
    \includegraphics[width=\textwidth]{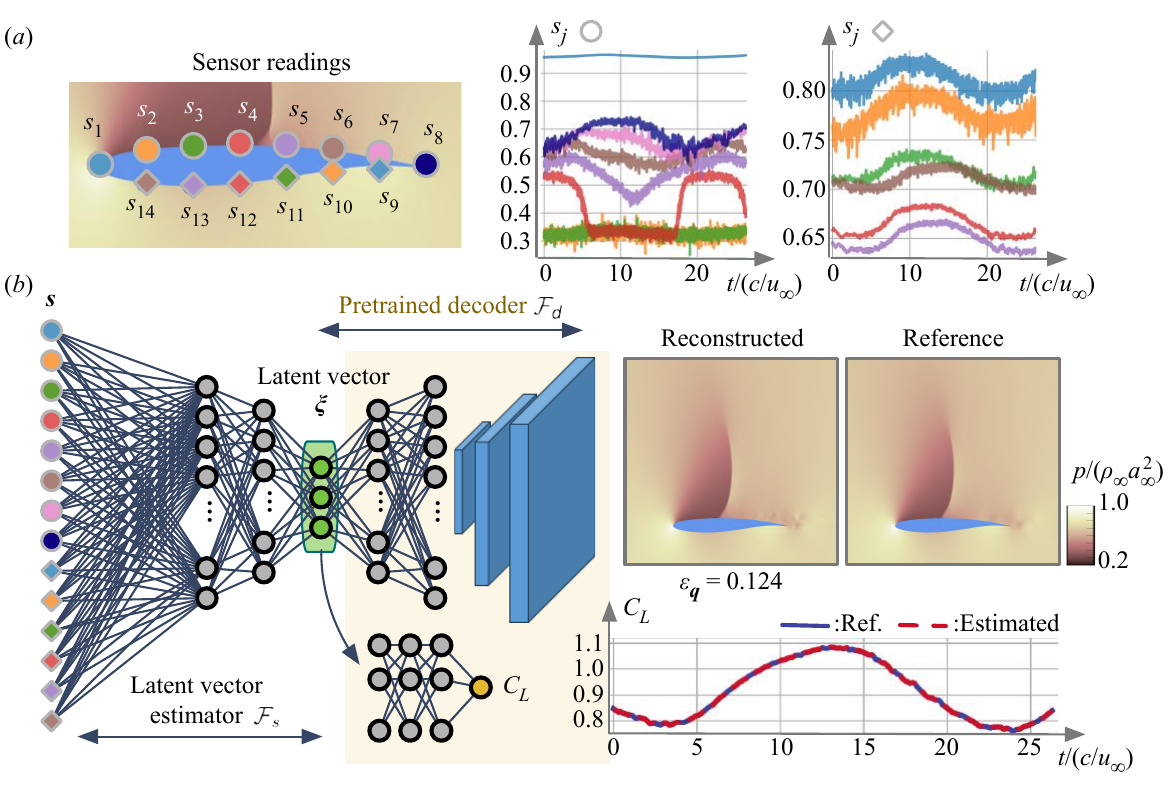}
    % \vspace{-5mm}
    \caption{
    Sparse sensor-based reconstruction via the low-order subspace.
    $(a)$ Pressure sensor placements on the wall and responses in time.
    $(b)$ The present full state reconstruction combined with a latent vector estimator {${\cal F}_s$} and the pretrained decoder ${\cal F}_d$.
    An example of the reconstructed field with the $L_2$ error norm $\varepsilon_{\bm q}$ and reproduced lift coefficient from fourteen sensors is shown.
    }
    % \vspace{-6mm}
    \label{fig7}
\end{figure}

The current findings above through the autoencoder compression imply that the right set of variables may capture the essence of transonic airfoil buffet flows.
This also makes us anticipate that sparse sensors could also be such a set of low-order variables, thereby achieving sparse-sensor-based reconstruction.
Furthermore, of interest here is whether it is possible to gain situational awareness from sparse sensors toward guiding flight operations based on insights into the physically-interpretable latent subspace. 
Based on this viewpoint, we further consider leveraging the discovered low-order subspace for the data-driven global flow field reconstruction.

Since the decoder ${\cal F}_d$ provides the pressure field from the latent vector, we aim to estimate the latent vector ${\bm \xi}(t)$ from sparse sensors ${\bm s}(t)$ by preparing an independent machine-learning model {${\cal F}_s$}. 
By feeding the estimated latent vector {$\hat{\bm \xi}={\cal F}_s({\bm s}(t))$} into the pretrained decoder~${\cal F}_d$, a pressure field ${\bm q}(t)$ is reconstructed, as illustrated in figure~\ref{fig7}.
The above-mentioned procedure is expressed as
{\begin{align}
    {\bm q}(t) \approx {\hat{\bm q}(t)} = {\cal F}_d(\hat{\bm \xi}(t)) = {\cal F}_d({\cal F}_s({\bm s}(t))),\label{eq:4-1}
\end{align}}
with an optimization for the weights {${\bm w}_s$} of the latent vector estimator {${\cal F}_s$},
{
\begin{align}
    {\bm w}_s^* = {\rm argmin}_{{\bm w}_s}||{\bm \xi} - {\cal F}_s({\bm s;{\bm w}_s})||_2^2.
\end{align}}
We use multi-layer perceptron~\cite{RHW1986} with the units of 14-32-64-128-32-3 across the layers for constructing the latent vector estimator {${\cal F}_s$} that maps sensor measurements ${\bm s}\in \mathbb{R}^{n_s}$ to ${\bm \xi} \in \mathbb{R}^3$, where $n_s$ represents the number of sensors.
This low-order mapping between sparse sensors and the latent vector enables avoiding a na\"ive learning for the relationship between the sensor inputs and the global field output~\cite{FFT2023_survey,EM2025}.
While such a field reconstruction problem often becomes computationally expensive due to a significant difference in data dimension between the input and output, this approach can save costs by leveraging the pretrained decoder.

\begin{figure}
    \centering
    % \hspace{-4mm}
    \includegraphics[width=1.05\textwidth]{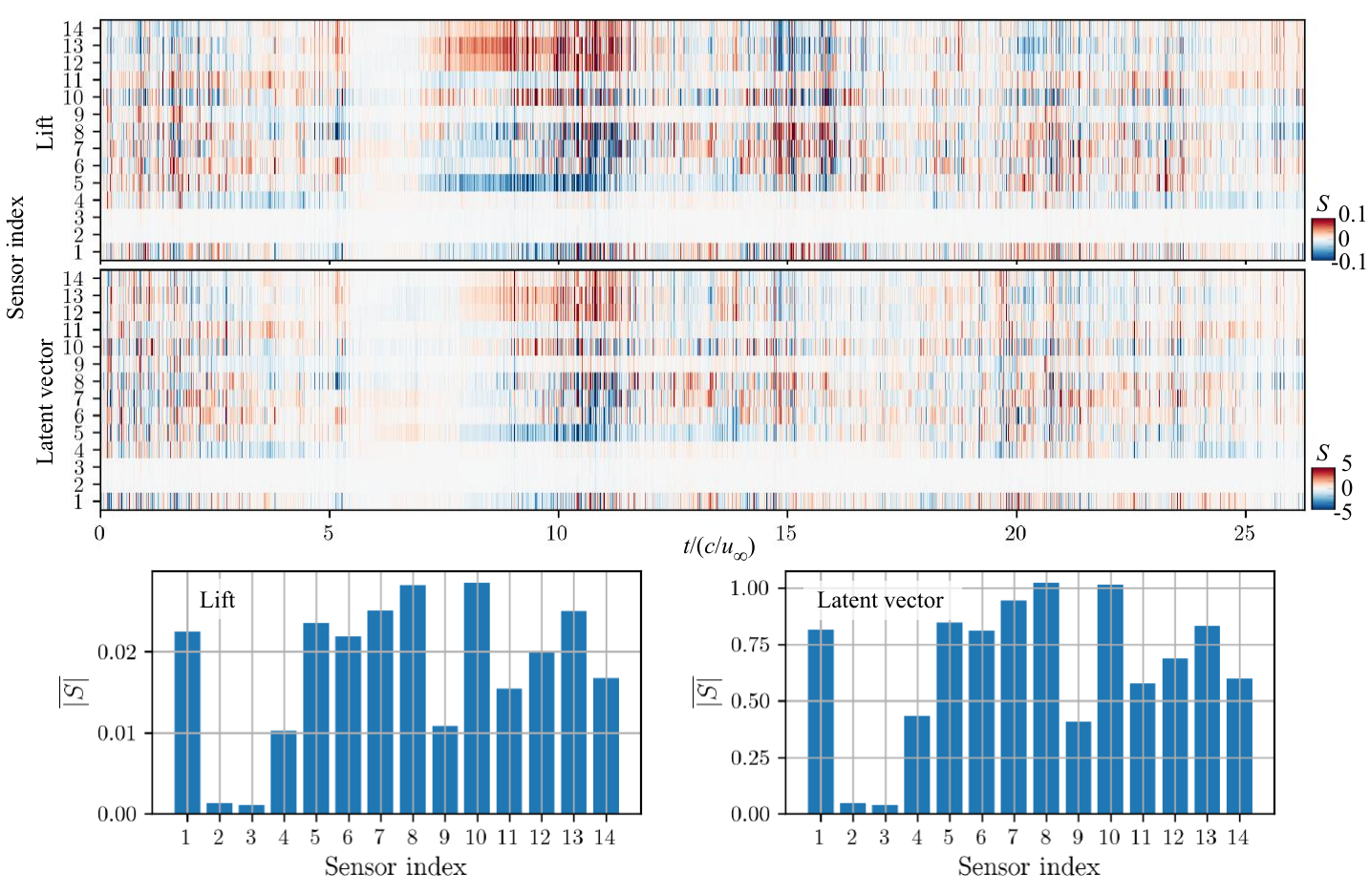}
    % \vspace{-5mm}
    \caption{
    Gradient-based sensor sensitivities with respect to the lift and latent vectors.
    Both the time trace and the time-averaged sensitivities over 14 sensors are shown.
    The sensor index here corresponds to shown in figures~\ref{fig7}, \ref{fig9} and~\ref{fig10}.
    }
    % \vspace{-6mm}
    \label{fig8}
\end{figure}

An example of the reconstructed pressure field and estimated lift coefficient from 14 sensors is shown in figure~\ref{fig7}.
Here, these sensors are placed along the airfoil surface in an equispaced manner{, enabling a comprehensive analysis of data-driven sensor reduction performed later.}
We use the latent vector ${\bm \xi}$ extracted from the lift-augmented autoencoder with $\beta = 0.05$.
In addition to the flow state including the wake shedding and shock location, the lift response is accurately reproduced from the sensor readings.
As implied through the discovery of a low-{dimensional} subspace, sparse sensor-based reconstruction is indeed possible for the present transonic airfoil buffet flow.

Furthermore, the minimal number and appropriate placements of sensors can be quantified with the latent vector estimator trained with 16 sensors above and the lift subnetwork prepared for subspace identification.
This is achieved by performing the sensitivity analysis between a machine-learning estimate and a given input~\cite{morimoto2022generalization,chen2024sparse}.
Considering the gradient between the sensor input ${\bm s}$ and the output of machine-learning model $\hat {\bm z}$, ${\bm \gamma}(t) = \partial {\hat {\bm z}}(t)/\partial {\bm s}(t)$, the importance of each sensor for estimation, i.e., sensitivity $S(t)$, is quantified as a weighted input,
{
\begin{align}
    S_j(t) = \gamma_j(t)s_j(t),
\end{align}}
where $j$ is an index of pressure sensor {$s_j$}.
As an output variable $\hat{\bm z}$, the estimated lift coefficient $\hat{C_L}$ and latent vector $\hat{\bm \xi}$ are considered.

\begin{figure}
    \centering
        % \hspace{-8.5mm}
    \includegraphics[width=0.8\textwidth]{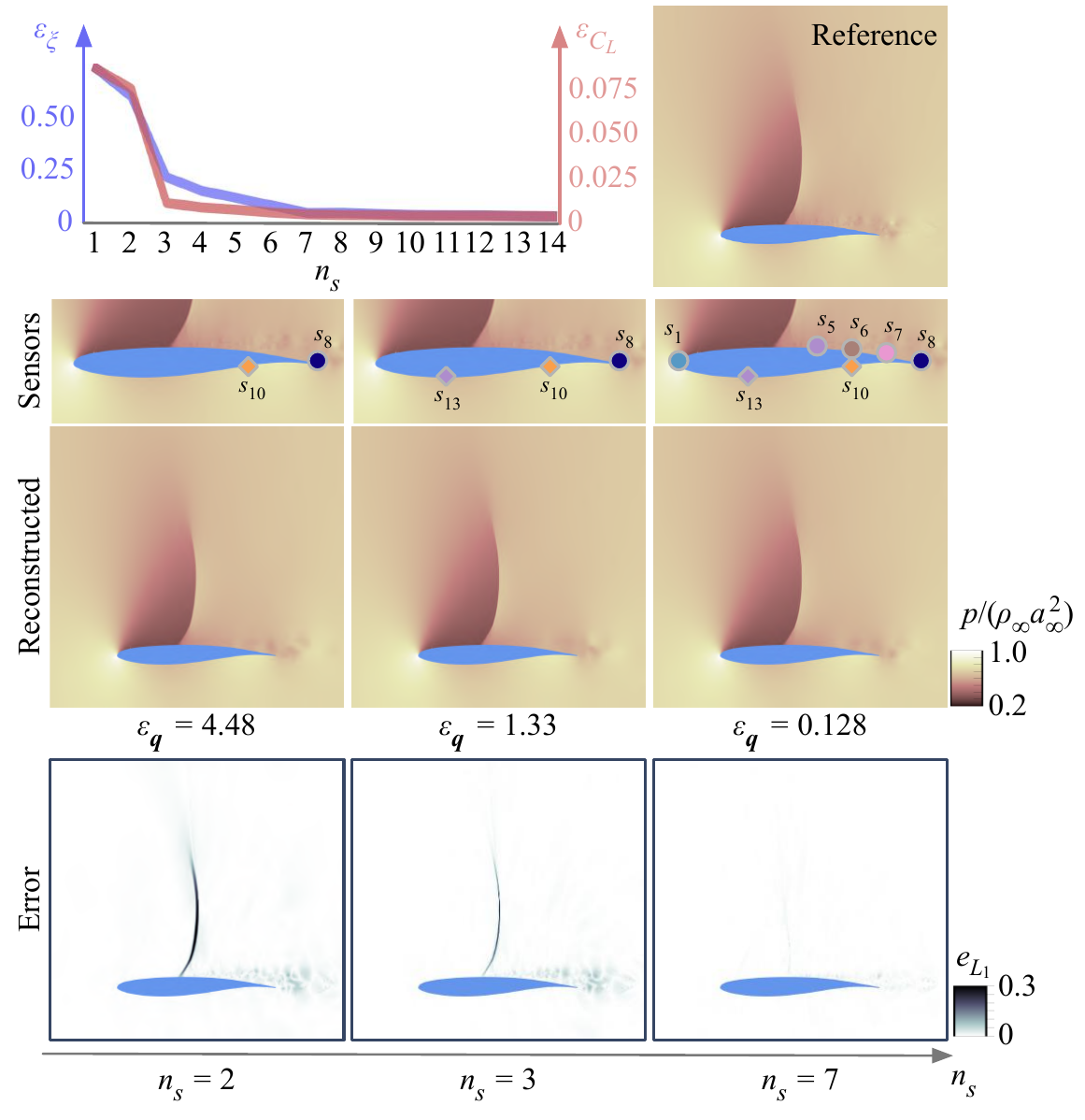}
    \vspace{-2mm}
    \caption{
    {
    Sensitivity-based sensor reduction.
    The relationship between the number of sensors $n_s$ and the estimation errors of latent vectors $\varepsilon_{\bm \xi}$ and lift ${\varepsilon}_{C_L}$ is shown.
    The reconstructed fields are presented with the $L_2$ error norm $\varepsilon_{\bm q}$ underneath each contour.
    }}
    \vspace{-5mm}
    \label{fig9}
\end{figure}

The sensitivity $S$ with respect to the lift and latent vectors is shown in figure~\ref{fig8}.
In addition to the time trace, the absolute time-averaged values are also presented to further gain insights into the general trend of sensitivities over the buffet cycle.
As the present autoencoder is trained such that the latent vector extracts the flow features associated with the lift coefficient, both sensitivity maps present a consistent trend in the direction of time and sensor index.
{Note that high-frequency fluctuations of the sensitivity are caused because the present sensitivity is calculated using the estimate by the machine-learned model, which includes the estimation error varying in time.
We have confirmed that the rank of sensor importance is not affected by such high-frequency fluctuations through a preliminary analysis by taking moving averages.
}

\begin{figure}
    \centering
        % \hspace{-8.5mm}
    \includegraphics[width=0.75\textwidth]{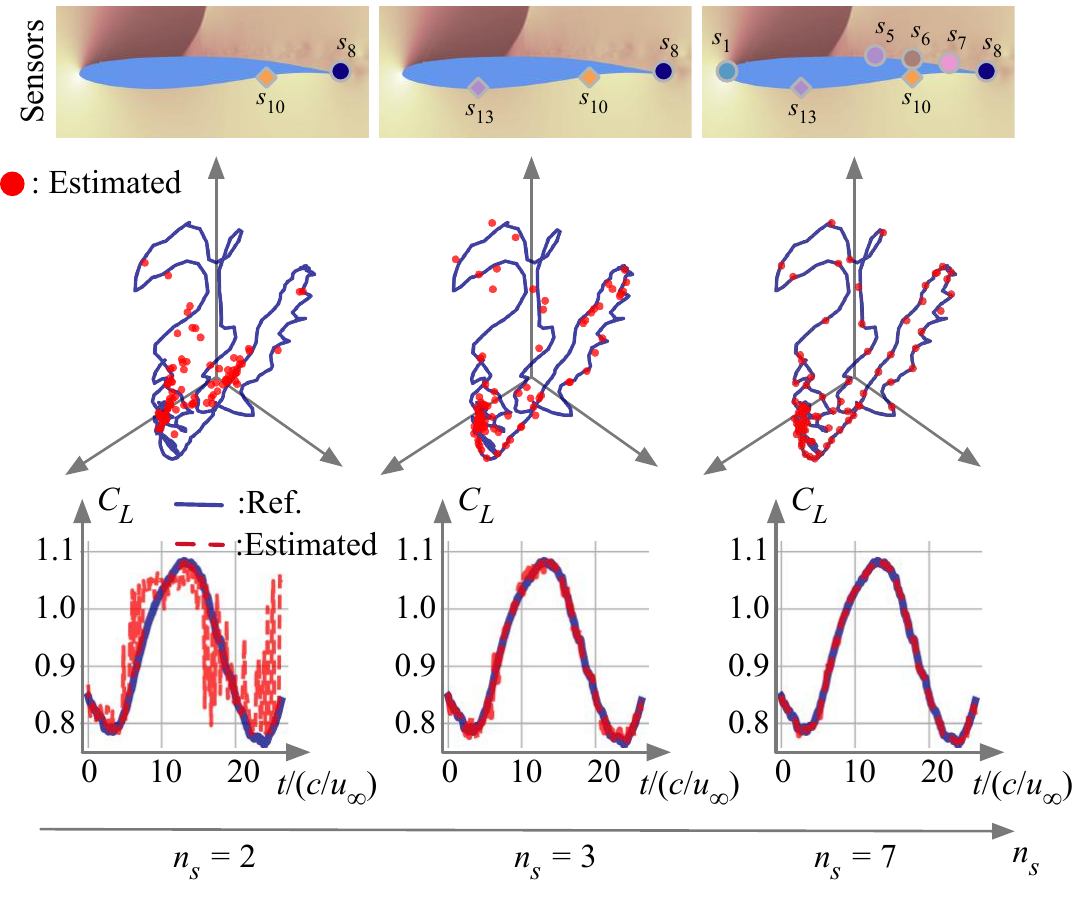}
    \vspace{-2mm}
    \caption{
    The estimated latent subspace and estimated lift coefficient across $n_s$ reduced via the sensitivity analysis.
    }
    \vspace{-3mm}
    \label{fig10}
\end{figure}

Focusing on the lift estimation, the sign of sensor sensitivity seems to be opposite between the suction (index 2-7) and pressure (index 9-14) sides due to their different role in contributing to lift. 
The responsible sensors are clearly shown where $\overline{|S|}>0.02$ --- sensor~1 at the leading edge, sensor 8 at the trailing edge, sensors 5, 6, and 7 placed on the suction side, and sensors 10 and 13 placed on the pressure side.
In turn, less sensitive sensors are also identified.
Sensors 2 and 3, placed in the supersonic region, particularly show very small $\overline{|S|}$, likely because their sensor signals are less affected by the shock movement compared to others according to figure~\ref{fig7}$(a)$.

The present sensitivity information is further leveraged to reduce the number of sensors for the subspace estimation.
Let us consider removing the sensors following the rank of absolute time-averaged sensitivity $\overline{|S|}$ so that sensors with small contribution to estimation are eliminated while keeping the highly contributing sensors.
The relationship between the number of sensors $n_s$ and the estimation errors is shown in figure~\ref{fig9}.
The error for the latent vector and lift response is depicted on a single plot.
The error curves are flat between $n_s=7$ and 14, exhibiting that accurate estimation of lift and flow fields is achieved up to $n_s=7$.
This is also evident from the reconstructed flow field shown in figure~\ref{fig9}, and the estimated latent subspace and lift response presented in figure~\ref{fig10}.

All seven sensor readings here report the absolute time-averaged value of $\overline{|S|}>0.02$, exhibiting a relatively larger value compared to other less-contributing sensors observed in figure~\ref{fig8}.
Once the sensors are further removed, the error of the latent vector starts to increase.
However, the error curve for the lift coefficient presents a slower slope at $n_{s}\leq 6$ compared to that for the latent space.
In fact, the lift response at $n_s = 3$ still exhibits reasonable agreement with the reference data.
This is likely because a global quantity of lift coefficient aggregating the flow information over the entire body is easier to estimate than the latent subspace, a representation of the whole flow field itself.

{To examine the dependence of reconstruction performance on the choice of sensor-selection technique and compression approach, we further consider the QR pivot-based sensor placement optimization \cite{manohar2018data} with $n_s = 7$.
Their approach finds the optimal sensor locations through the QR factorization with column pivoting applied to the POD bases.
Further details on this linear technique are referred to Manohar et al.~\cite{manohar2018data}.
The original placements of sensors before performing the QR pivot-based reduction are constrained on the wing surface and the same as those used in the autoencoder-based analysis shown in figure~\ref{fig7}$(a)$.
Here, three approaches are considered, namely;
\begin{enumerate}
    \item Estimate the three-dimensional latent vectors ${\bm \xi}$ based on the sensors reduced via the gradient sensitivity and decode a flow using the nonlinear decoder ${\cal F}_d$ (the original formulation)
    \item Estimate the dominant three POD coefficients ${\bm a}$ based on the sensors reduced via the QR pivot and decode a flow with POD modes
    \item Estimate the three-dimensional latent vectors ${\bm \xi}$ based on the sensors reduced via the QR pivot and decode a flow using the nonlinear decoder ${\cal F}_d$ 
\end{enumerate}
For fair comparison, we use the same MLP architecture for all three cases in estimating the latent vectors and the three dominant POD coefficients.
The flow fields are then decoded using the nonlinear decoder ${\cal F}_d$ or POD modes ${\bm \Phi}$.
While equation~\ref{eq:4-1} is applied for cases (i) and (iii), case (ii) with the POD-MLP model with the QR pivot-based sensor reduction is expressed as 
\begin{align}
    {\bm q}(t) \approx  {\bm \Phi}\hat{\bm a}(t) = {\bm \Phi}({\cal F}_{s}({\bm s}(t))).\label{eq:4-QR}
\end{align}

\begin{figure}
    \centering
    \includegraphics[width=0.75\textwidth]{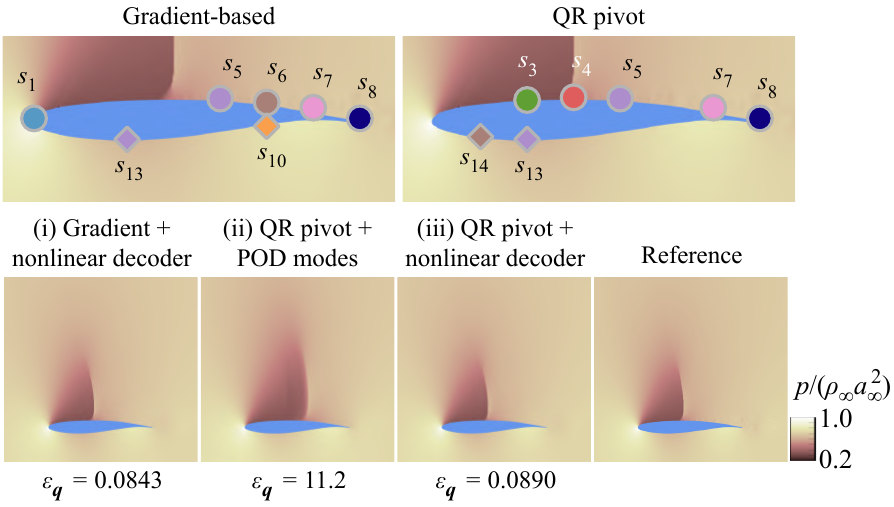}
    % \vspace{-5mm}
    \caption{
    {
    Dependence of sparse-sensor reconstruction performance on the choice of sensor-reduction technique and compression approach with $n_{\bm \xi} = 7$.
    }
    }
    \vspace{-3mm}
    \label{fig_sens}
\end{figure}

Let us compare the reduced sensor placements in figure~\ref{fig_sens}.
Four sensors (index 5, 7, 8, and 13), reporting high $\overline{|S|}$ with the gradient-based approach, are commonly kept with both sensor-reduction methods through the reduction process.
However, the remaining three sensors are placed in a different way.
While sensors chosen by the QR pivot are grouped with neighbors ($s_3$--$s_5$, $s_7$--$s_8$, and $s_{13}$--$s_{14}$), the gradient-based method seems to attempt to cover the entire wing surface. 
This result suggests that the dominant features captured by both POD and the autoencoder make the reduction approach keep the common four sensors, while the subdominant characteristics that are better compressed with the nonlinear autoencoder cause the difference in the location of the remaining three sensors.

The reconstruction fields with cases (i-iii) are also shown in figure~\ref{fig_sens}.
When using the nonlinear decoder, the reconstruction with the gradient-based approach is slightly better than that with the QR pivot.
These accurate reconstructions suggest that the high error for case (ii) is primarily due to the use of linear POD modes as a decoder rather than the sensor placements determined by the QR pivot.
We note that the error for case (iii) of the QR pivot and the autoencoder latent variables starts to increase with $n_s \leq 6$, similarly to case (i) using the gradient-based method, although not shown.
While both the gradient-based method and the QR pivot currently provide a similar level of sensor reduction performance, they could be further improved by accounting for redundancy between sensor readings, which can be quantified with inter-correlations and mutual information.
}

The present analysis above is focused on the transonic airfoil buffet flow at $Re = 3\times 10^6$.
While the current Reynolds number may be higher than those often considered for numerical and data-driven analyses in the community, this still resides in a range of wind-tunnel scale conditions. 
Of particular interest here is whether the current model trained at a wind-tunnel scale Reynolds number can be applied to a scenario under the real aircraft operation level of Reynolds number.
In response, this study lastly evaluates the applicability of the present method to a transonic airfoil buffet flow at $Re=3\times 10^7$ with $M_\infty = 0.730$.

\begin{figure}
    \centering
    \includegraphics[width=\textwidth]{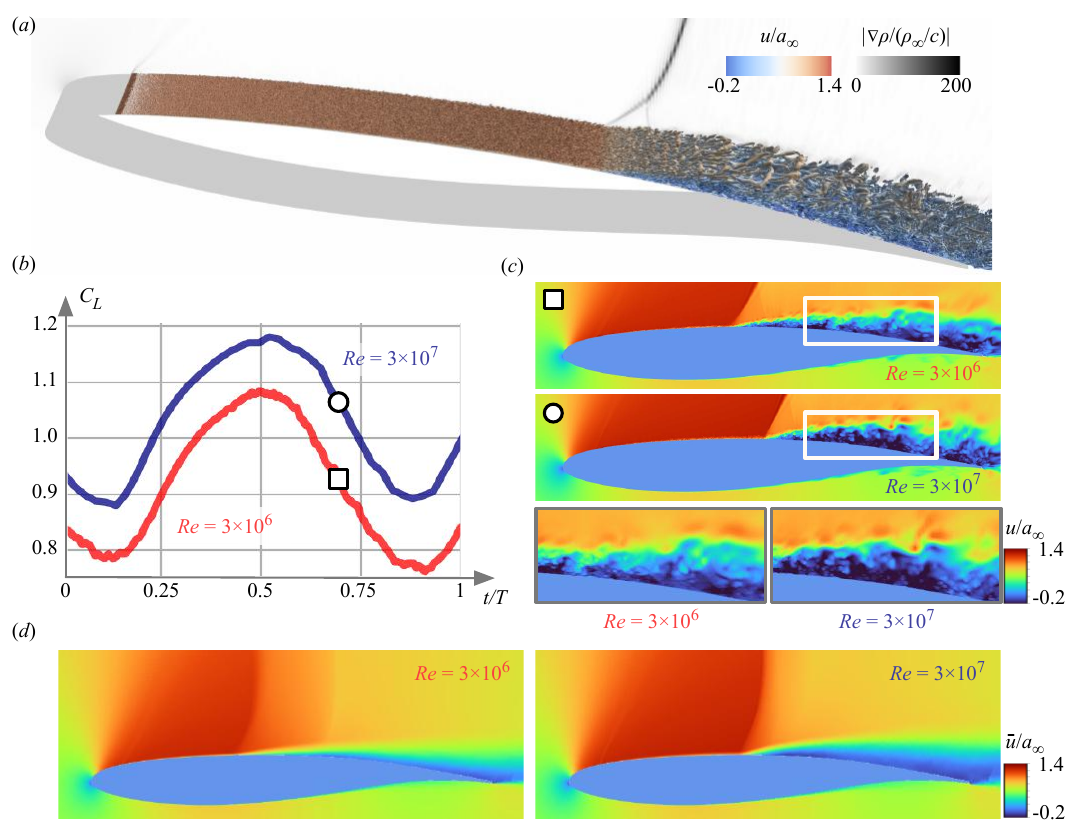}
    % \vspace{-5mm}
    \caption{
    $(a)$ An instantaneous snapshot of transonic airfoil buffet flows at $Re = 3\times 10^7$ visualized by the isocontours of $Q$-criterion.
    Comparison of 
    $(b)$ lift coefficient and $(c)$~instantaneous streamwise velocity fields sampled at $t/T = 0.70$ with $Re=3\times 10^6$ and $Re=3\times 10^7$.
    {$(d)$~Time- and spanwise-averaged streamwise velocity fields at $Re=3\times 10^6$ and $Re=3\times 10^7$.}
    }
    \vspace{-3mm}
    \label{fig11}
\end{figure}

The wall-modeled LES is performed for the case with $(Re,M_\infty) = (3\times 10^7, 0.730)$ at $\alpha = 3.5^\circ$, as presented in figure~\ref{fig11}$(a)$.
There is a self-sustained shock buffet cycle {that produces} almost the same frequency and oscillation amplitude of aerodynamic coefficients as that for $Re=3\times 10^6$, as seen in figure~\ref{fig11}$(b)$.
The difference in the flow between two Reynolds numbers is examined with the instantaneous streamwise velocity $u$ sampled at the same phase $t/T=0.70$ where $T$ denotes the time window across the buffet cycle, as depicted in figure~\ref{fig11}$(c)$.
The shock location moves downward {and the separation height becomes higher} by increasing the Reynolds number, strengthening the shock wave accompanied by a large adverse pressure gradient and triggering the larger separation{, which is also evident from the time-averaged flow fields shown in figure~\ref{fig11}$(d)$}.
Due to the trade-off relationship between the suppression effect of separation due to the increment of Reynolds number and the separation induced by the strong shock wave, the resulting shock wave oscillation is sustained.

\begin{figure}[t]
    \centering
    \includegraphics[width=\textwidth]{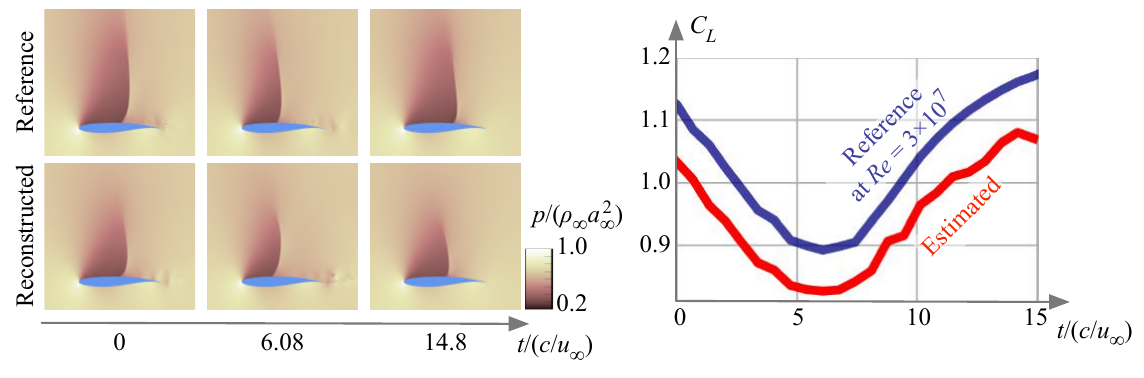}
    \vspace{-6mm}
    \caption{
    Application of the sparse sensor reconstruction model trained at $Re=3\times 10^6$ to a flow at the level of a real aircraft operation of $Re=3\times 10^7$.
    The reconstructed pressure field and lift response are shown. 
    }
    \vspace{-3mm}
    \label{fig12}
\end{figure}

Let us finally apply the present sensor-based reconstruction model trained at $Re = 3\times 10^6$ to the level of real-aircraft operation at $Re = 3 \times 10^7$, as shown in figure~\ref{fig12}.
Here, we use the latent vector estimator ${\cal F}_p$ trained with seven sensors following the observation in figures~\ref{fig9} and \ref{fig10}.
The reconstructed fields exhibit a smaller height of shock compared to the reference snapshots as such a shock with a higher height does not appear in the training data at $Re=3\times 10^6$.
However, it is worth noting that the shock locations of the machine-learning reconstruction are constantly evaluated forward compared to that of the reference at $Re = 3\times 10^7$.
Since the shock moves downward by increasing the Reynolds number while keeping its phase as presented in figure~\ref{fig11}, this constant shift indicates that the present model may correctly capture the phase information across the buffet cycle even at the current real aircraft-level Reynolds number.
This is further evident from the reproduced lift response.
While the magnitude of lift is underestimated due to the difference in Reynolds number between the training and testing data, the temporal trend of the lift signal accurately matches the reference.
This observation suggests that nonlinear machine learning can {be transferred to scenarios where the characteristics of variables of interest remain relatively consistent across different Reynolds numbers.}

% \vspace{-4mm}
\section{Concluding remarks}
\label{sec:conc}

This study examined a low-{dimensional} representation of transonic airfoil buffet flows at a high Reynolds number with nonlinear machine learning. 
Wall-modeled large-eddy simulations of flow over the OAT15A supercritical airfoil at Mach numbers $M_\infty = 0.715$ and 0.730, corresponding to non-buffet and buffet conditions, were performed at a chord-based Reynolds number of $Re = 3\times 10^6$ to generate the data sets used in the present data-driven analysis.
To derive a low-order expression from the data, we considered nonlinear lift-augmented autoencoder-based compression.
We found that there exists a compact three-dimensional latent subspace reflecting the characteristics of transonic airfoil buffet flow.
The discovered representation captures key flow features, including shock movement and shock-induced separation, in a reduced-order manner.

Based on these physical implications, sparse sensor-based reconstruction via the learned representation was further performed.
Equipped with the sensitivity analysis, the sensor configuration required for accurately reproducing aerodynamic responses can be determined.
Finally, the model trained at a wind tunnel scale Reynolds number of $Re = 3\times 10^6 $ was assessed at a real aircraft operational level of $Re = 3\times 10^7$, revealing its ability to reasonably predict phase dynamics of aerodynamic loads from sparse sensors.

{
While we considered two configurations of buffet/non-buffet conditions at a fixed angle of attack, additional cases with a range of different parameters, including angle of attack, Reynolds number, and Mach number, would be needed to fully characterize the whole picture of buffet onset.
Although it is anticipated that a low-order subspace capturing the difference in such parameters and the occurrence of transonic buffet could be identified, a major challenge arises from a collection of data sets through large-scale simulations.
From this aspect, one can consider data fusion between LES, unsteady RANS, and experimental measurements to supplement the pros and cons across different data sets with each other in extracting a low-order submanifold with observable-augmented learning~\cite{fukami2025observable}.
}

{
The present analysis reveals that three latent variables are needed to represent transonic airfoil buffet flows. 
Although buffet dynamics are often modeled as a self-sustained oscillator subjected to stochastic forcing~\cite{feldhusen2021analysis,sansica2022system,crouch2024weakly}, our findings suggest the necessity of a third dimension.
This additional latent dimension likely corresponds to aerodynamic phenomena related to the separation height, according to the observation in figure~\ref{fig6}. 
To characterize these dynamics more precisely, it is essential to investigate the nonlinear modal structures associated with each latent variable. 
This can be achieved by integrating mode-decomposing autoencoders~\cite{MFF2019,FNF2020}, which we plan to pursue in future work.
}

{
With the present formulation of observable-augmented learning, users have to choose an appropriate observable from the candidates, and it currently takes some level of computational effort to find a physically-relevant subspace.
Note, however, that the former point of the non-automatic process enables us to have the opportunity to incorporate physical or mathematical knowledge based on what we would like to associate with, while the computational cost for the latter point is still manageable as the degree of freedom of observables is much less than that of the original simulations.
A series of recent studies on observable-augmented manifold learning have revealed that an appropriate choice of observable assists in compactly extracting physics for a range of unsteady flow scenarios including vortex-airfoil interactions~\cite{FT2023,fukami2024data,liu2024model,ME2025}, vehicle aerodynamics~\cite{tran2024data}, turbulent boundary layers~\cite{fukami2025observable}, and roughness turbulence~\cite{nair2025rough}, enabling the enjoyment to learn physics from data for fluid mechanicians.
More broadly, an ``observable" here does not need to be a variable.
Some applied mathematical techniques, such as persistent homology~\cite{smith2024cyclic} and information theory~\cite{fukami2025information}, can also be considered as observables depending on the physics of interest.
Hence, adding an observable may be regarded as one approach to support data-driven analysis for unsteady flows.
}

Based on the current findings considering flows around a wing, the applicability of the present data-driven subspace identification to transonic buffet conditions around a full-aircraft configuration would also be of interest~\cite{asada2023ffvhc,tamaki2024wall}.
{For such cases, a combination of linear, scalable compression techniques such as POD and the present observable augmentation would be helpful to reduce the computational burden~\cite{linot2023dynamics,tran2024data,asada2025exact}.}
The current study may offer a new perspective on the analysis and determination of flight envelope toward next-generation air vehicle operations.

\vspace{-3mm}
\section*{{Appendix A: Training procedures and L-curve analysis}}
\label{sec:App1}

\begin{figure}[b]
    \centering
    \includegraphics[width=0.75\textwidth]{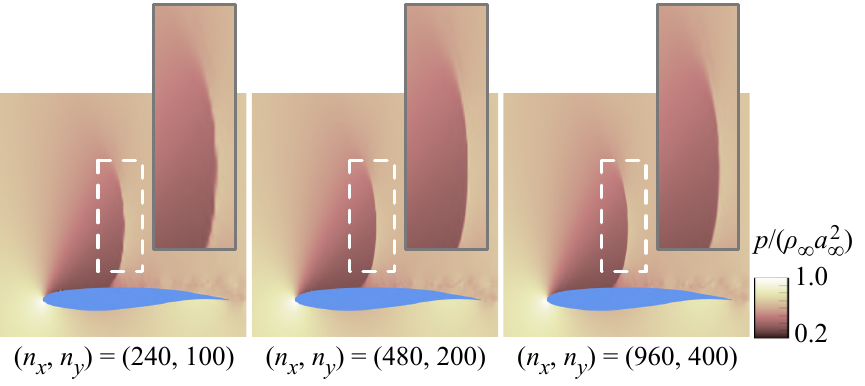}
    % \vspace{-6mm}
    \caption{
    {
    Pressure field interpolated onto the spatially uniform grid with the resolution of $(n_x,n_y) = (240,100)$, $(480,200)$, and $(960,400)$.
    }
    }
    % \vspace{-3mm}
    \label{fig_App0}
\end{figure}

\begin{figure}
    \centering
    \includegraphics[width=0.6\textwidth]{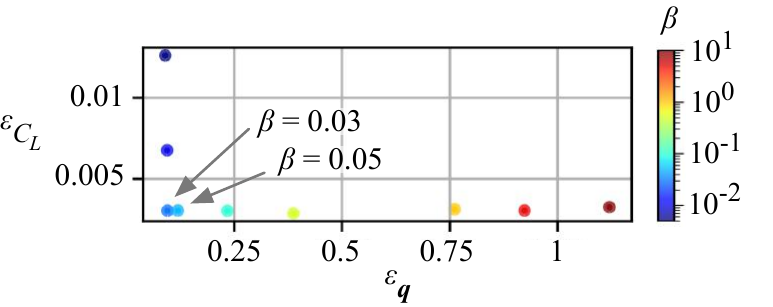}
    % \vspace{-6mm}
    \caption{
    {
    L-curve analysis for the present observable-augmented autoencoder.
    }
    }
    \vspace{-3mm}
    \label{fig_App1}
\end{figure}

{
Here, we provide details on training procedures and the choice of weighting parameter $\beta$ in equation~\ref{eq:obsloss} for the present observable-augmented nonlinear autoencoder.
The Adam optimizer~\cite{Kingma2014} with the default parameter sets in Keras is used to update the weights through the machine-learning training.
The maximum number of training iterations is set to be 50,000, while early stopping~\cite{prechelt1998automatic} with the criterion of a series of continuous 100 epochs is employed to avoid overfitting.
We use 70\% of the data sets for training and the remaining 30\% are prepared for validation.
The number of grid points $(N_x,N_y) = (480,200)$ for the current data-driven analysis is determined such that the shock can be represented without exhibiting any discontinuous artifacts, which is evident from the comparison to other resolutions $(N_x,N_y) = (240,100)$ and $(960,400)$ shown in figure~\ref{fig_App0}. 
In using the entire data set of 24,100 snapshots, the training process takes approximately two hours in an NVIDIA A100 GPU environment, and the inference time for each snapshot is 0.003 seconds.

The weighting parameter $\beta$ in equation~\ref{eq:obsloss} is determined based on the L-curve analysis~\cite{hansen1993use} that finds an appropriate regularization parameter of the cost function, as shown in figure~\ref{fig_App1}.
We consider nine different $\beta$ (0.005, 0.01, 0.03, 0.05, 0.1, 0.5, 1, 5, and 10).
The cases with $\beta =0.03$ and 0.05, providing low reconstruction errors for the lift response and the pressure field in a balanced manner, are chosen for the present analysis.  
}

% \vspace{-4mm}
\section*{{Appendix B: Reconstructed variables for the non-buffet case}}
\label{sec:App2}

{

\begin{figure}[t]
    \centering
    \includegraphics[width=0.85\textwidth]{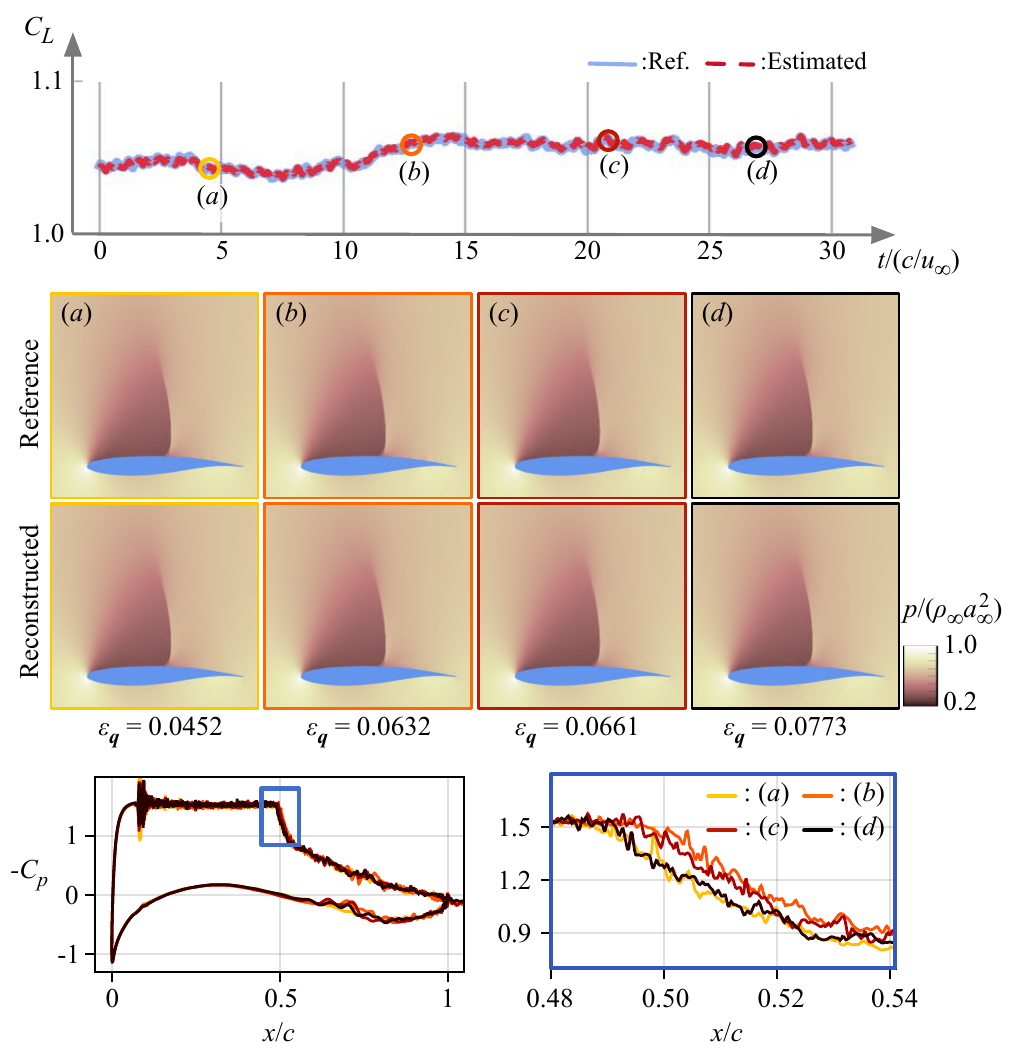}
    % \vspace{-6mm}
    \caption{
    {
    Decoded lift coefficient and pressure fields via a lift-augmented autoencoder with $\beta = 0.05$ for the non-buffet case with $M_\infty = 0.715$.
    The flow fields $(a)-(d)$ correspond to shown in figure~\ref{fig2}.
    The whole (left) and zoom-in (right) views of the pressure coefficient $C_p$ on the wing surface for the snapshots $(a)-(d)$ are also presented.
    }
    }
    \vspace{-3mm}
    \label{fig_App2}
\end{figure}

Let us exhibit in figure~\ref{fig_App2} the decoded lift coefficient and pressure fields obtained from the present lift-augmented autoencoder with $\beta = 0.05$.
While achieving accurate estimation of lift coefficient, the reconstructed pressure fields are in agreement with the reference data, reporting less than 8\% $L_2$ norm error over time.
Along with the observation of a small-sized cyclic orbit in figure~\ref{fig5} and small oscillations of the pressure coefficient $C_p$ in figure~\ref{fig_App2}, it is argued that the present model represents statistically steady dynamics of the non-buffet case well in the identified low-order subspace.}

% \vspace{-3mm}
\section*{{Appendix C: Effect of the number of training samples}}
\label{sec:App3}

{

We examine the dependence of reconstruction performance and latent space geometry on the number of training snapshots by subsampling them to be 25\% and 50\% of the original amount, as presented in figure~\ref{fig_App3}.
We use the same autoencoder network with the same weighting parameter $\beta$ of 0.05 as that used in the original case.
The case with 50\% presents a similar result to the original model.
However, the latent geometry with the 25\% case starts to deform from the original shape, although there still exists a two-wing-shaped submanifold.
Since the dimensionality in the subspace is determined based on whether the given data covers the entire space of the attractor or not, rather than the number of snapshots, the latent dimension is not affected for this analysis, in which we subsample the snapshots while keeping the entire time window.

The deformation of latent space geometry is caused by several reasons.
There may exist an optimal weighting parameter $\beta$ for the case with 25\% data.
Furthermore, the primary reason is likely less temporal density of data compared to the original case, which may cause miscapturing some events over the buffet cycle. 
A sufficient temporal resolution is needed to obtain an interpretable low-order subspace in a data-driven manner.
}

% \vspace{-5mm}
\section*{{Appendix D: Uniqueness of latent representation}}
\label{sec:App4}

{

\begin{figure}[t]
    \centering
    \includegraphics[width=0.9\textwidth]{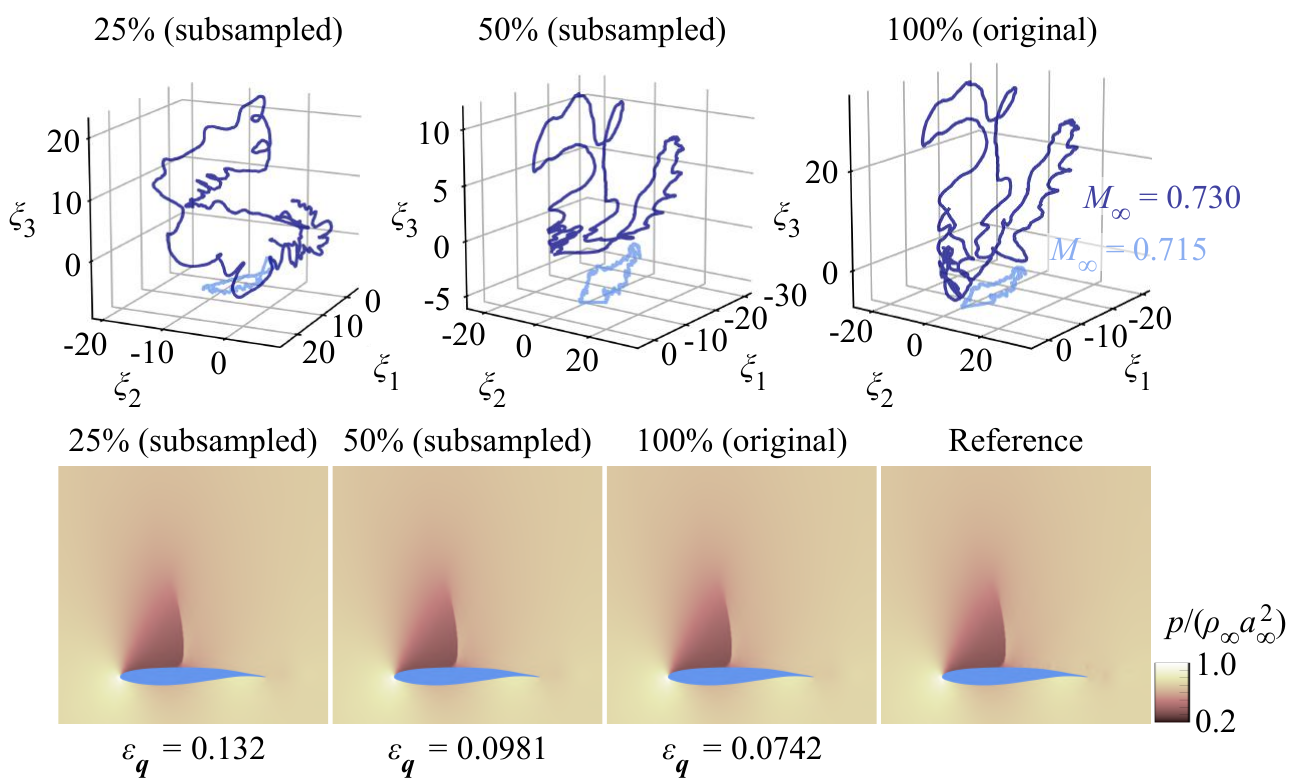}
    % \vspace{-6mm}
    \caption{
    {
    Dependence of field reconstruction performance and latent space geometry on the number of training snapshots.
    }
    }
    % \vspace{-3mm}
    \label{fig_App3}
\end{figure}

\begin{figure}
    \centering
    \includegraphics[width=0.9\textwidth]{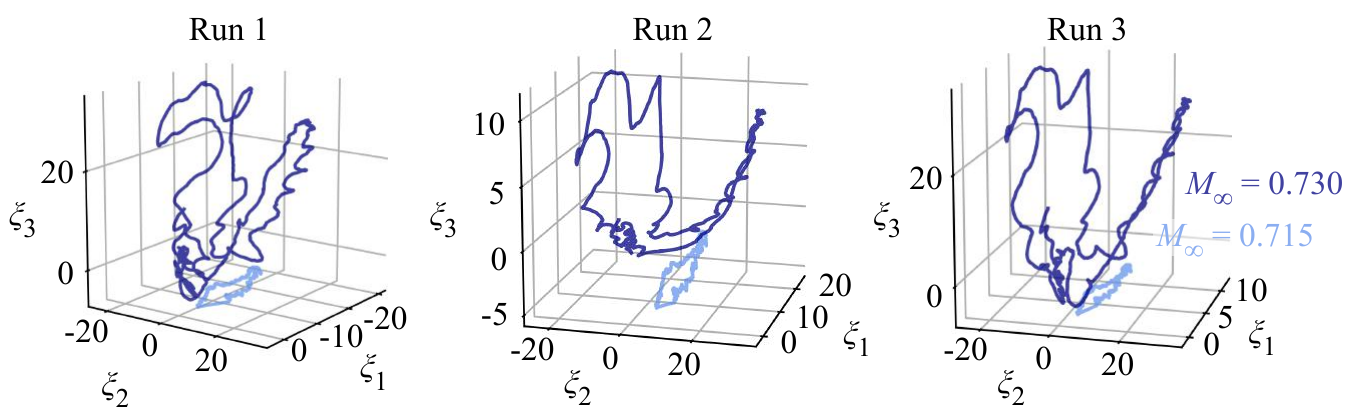}
    % \vspace{-6mm}
    \caption{
    {
    Dependence of the latent geometry on the initial random seed assigned to the weights in the observable-augmented autoencoder.
    }
    }
    % \vspace{-3mm}
    \label{fig_App4}
\end{figure}

To consider the uniqueness of latent representation, we examine the dependence of the latent geometry on the initial random seed assigned to the weights in the observable-augmented autoencoder, as shown in figure~\ref{fig_App4}.
The weighting parameter $\beta$ of 0.05 is used for this analysis.
The model exhibits reasonable robustness across the three runs, presenting a two-wing-shaped submanifold while distinguishing the non-buffet and buffet cases in a low-order manner. 
Although this paper only considers a single network configuration of observable-augmented autoencoder, the results above indicate that a model may provide a similar wing-shaped geometry over a range of the network capacities by choosing the optimal value of $\beta$ through the L-curve analysis.
}

%%%%%%%%%%%%%%%%%%%%%%%%%%%%%%%%
%%%%%%%%%%%%%%%%%%%%%%%%%%%%%%%%%
%%%%%%%%%%%%%%%%%%%%%%%%%%%%%%%%%

\bibliographystyle{unsrt}  
\bibliography{refs}

\end{document}